\documentclass[11pt,a4paper]{article}
\usepackage{epsfig}
\usepackage[T1]{fontenc}    
\usepackage{graphics}
\usepackage{graphicx}
\usepackage{pstricks,pst-coil,pst-fill,pst-plot}
\usepackage[fleqn]{amsmath}    
\usepackage{amssymb}    
\usepackage{amsfonts}   
\usepackage{verbatim}   
\usepackage{mathrsfs}   
\usepackage{dsfont}
\usepackage{euscript}
\usepackage{yfonts}
\usepackage{enumerate}     
\usepackage{amsthm}         
\usepackage{txfonts}
\usepackage{marvosym}
\usepackage{stmaryrd}
\usepackage{vmargin}        
\usepackage{wasysym}		

\setmarginsrb{1.8cm}{2cm}{1.8cm}{2cm}{1cm}{1cm}{1cm}{1.6cm}
 \makeatletter
 \@addtoreset{equation}{section}
 \makeatother

\providecommand{\bysame}{\leavevmode\hbox to3em{\hrulefill}\thinspace}
\providecommand{\MR}{\relax\ifhmode\unskip\space\fi MR }

\providecommand{\href}[2]{#2}

\let\ua=\uparrow
\let\da=\downarrow
\let\tend=\rightarrow

\long\def\symbolfootnote[#1]#2{\begingroup%
\def\thefootnote{\fnsymbol{footnote}}\footnote[#1]{#2}\endgroup}

\newtheorem{theorem}{Theorem}[section]

\newtheorem*{theorem*}{Theorem}

\newtheorem{conj}[theorem]{Conjecture}

\newcommand\beq{\begin{equation}}
\newcommand\enq{\end{equation}}
\newcommand\bem{\begin{multline}}
\newcommand\enm{\end{multline}}

\def\beqa{\begin{eqnarray}}
\def\eeqa{\end{eqnarray}}
\def\ba{\begin{array}}
\def\ea{\end{array}}
\def\det{\operatorname{det}}

\newcommand{\f}[2]{{\ensuremath{%
    \mathchoice%
    {\dfrac{#1}{#2}}
    {\dfrac{#1}{#2}}
    {\frac{#1}{#2}}
    {\frac{#1}{#2}}
}}}
\newcommand{\tf}[2]{\ensuremath{#1/#2}}
\def\a{\alpha}

\def\Ga{\Gamma}

\def\de{\delta}

\def\De{\Delta}
\def\eps{\epsilon}
\def\veps{\varepsilon}
\def\la{\lambda}
\def\La{\Lambda}

\def\sg{\sigma}

\def\ups{\upsilon}
\def\th{\theta}

\def\om{\omega}

\newcommand{\mc}[1]{\ensuremath{\mathcal{#1}}}
\newcommand{\mf}[1]{\ensuremath{\mathfrak{#1}}}
\newcommand{\msc}[1]{\ensuremath{\mathscr{#1}}}

\newcommand{\bs}[1]{\ensuremath{\boldsymbol{#1}}}

\DeclareFontFamily{OT1}{pzc}{}
\DeclareFontShape{OT1}{pzc}{m}{it}{<-> s * [1.10] pzcmi7t}{}
\DeclareMathAlphabet{\mathpzc}{OT1}{pzc}{m}{it}

\def \i{ \mathrm i}

\newcommand{\wt}[1]{\ensuremath{\widetilde{#1}}}
\newcommand{\wh}[1]{\ensuremath{\widehat{#1}}}

\newcommand{\Int}[2]{\ensuremath{\int\limits_{#1}^{#2}}}
\newcommand{\Oint}[2]{\ensuremath{\oint\limits_{#1}^{#2}}}

\newcommand{\sul}[2]{\ensuremath{\sum\limits_{#1}^{#2}}}
\newcommand{\pl}[2]{\ensuremath{\prod\limits_{#1}^{#2}}}

\newcommand{\R}{\ensuremath{\mathbb{R}}}
\newcommand{\Cx}{\ensuremath{\mathbb{C}}}

\newcommand{\Dp}[1]{\ensuremath{\partial_{#1}}}

\newcommand{\ex}[1]{\ensuremath{\e{e}^{#1}}}

\newcommand{\op}[1]{ \boldsymbol{ \texttt{#1} } }

\newcommand{\dd}{\mathrm{d}}
\newcommand{\e}[1]{\ensuremath{\mathrm{#1}}}

\newcommand{\intff}[2]{\ensuremath{ [  #1 \,; #2 ] }}

\newcommand{\intoo}[2]{\ensuremath{ ]  #1 \,; #2 [ }}

\newcommand{\intn}[2]{\ensuremath{[\![ \, #1 \,;\, #2 \,]\!]}}

\begin{document}

\begin{center}
\begin{LARGE}
{\bf Space-like asymptotics of the thermal two-point functions of the XXZ spin-1/2 chain}
\end{LARGE}

\vspace{1cm}

{\large
Frank G\"{o}hmann\symbolfootnote[1]{e-mail: goehmann@uni-wuppertal.de}}
\\[1ex]
Fakult\"at f\"ur Mathematik und Naturwissenschaften, Bergische Universit\"at
Wuppertal, 42097 Wuppertal, Germany.\\[2.5ex]

{\large Karol K. Kozlowski\symbolfootnote[2]{e-mail: karol.kozlowski@ens-lyon.fr}}
\\[1ex]
ENSL, CNRS, Laboratoire de physique, F-69342 Lyon, France. \\[2.5ex]

\par 

\end{center}

\vspace{40pt}

\centerline{\bf Abstract} \vspace{1cm}
\parbox{12cm}{\small}

This work proposes a closed formula for the leading term of  the long-distance and large-time asymptotics in a cone of the space-like regime for the
transverse dynamical two-point functions of the XXZ spin 1/2 chain at finite temperatures. The result follows from a simple
analysis of the thermal form factor series for dynamical correlation functions. The leading asymptotics we obtain are driven by the Bethe Ansatz data
associated with the first sub-leading Eigenvalue of the quantum transfer matrix.

\vspace{40pt}

\tableofcontents

\section{Introduction}

Dynamical correlation functions constitute the fundamental observables of many-body quantum systems in that their Fourier transforms
are directly measured in experiments. Despite their central role in connecting experiments with theory, their full theoretical understanding
is still in its infancy. This can be mostly ascribed to their mathematical intricacy which makes such observables extremely hard to compute directly.
In arbitrary dimension, dynamical correlation functions may be studied by means of numerical techniques
such as Monte Carlo simulations, exact diagonalisation or tensor network methods.
These techniques rely on various \textit{ad hoc} approximations or simplifications which are hard to test due to the lack of
exact results. Moreover, while quite effective for static correlators, such techniques loose accuracy for dynamical correlators,
especially in the long-time, large-distance regime. Yet, this regime is interesting for various reasons. First of all, one expects that -at least for specific values of
coupling constants and temperatures- there should emerge a universal behaviour. Second, this asymptotic regime leads to simplifications in the expressions
for the correlators which may help to provide, \textit{a posteriori}, once that they are obtained, good phenomenological models
for dynamical correlation functions which may allow one to access the overall qualitative behaviour of the latter.

The situation is quite special in one spatial dimension owing to the existence of numerous exactly solvable models having  genuine non-trivial interactions
and which capture the essence of real compounds, \textit{viz}. are  relevant to experiments.
For these models, one may access  closed explicit formulae for the Eigenvectors and Eigenvalues by means of some variant of the
Bethe Ansatz \cite{BetheSolutionToXXX}. One may build on the quantum inverse scattering method so as to compute their correlation functions, be it static or dynamic.
The attention in devising techniques allowing one to do so  was mostly focused -especially in what concerns correlation functions-
on the XXZ spin-1/2 chain which is a paradigmatic example
of a quantum integrable model. It is described by the Hamiltonian operator subject to periodic boundary conditions, \textit{i.e.}
$\sg_{L+1}^{\a}=\sg_{1}^{\a}$:
\beq
\op{H} \, = \, J \sum_{a=1}^{L} \Big\{ \sigma_a^x \,\sigma_{a+1}^{x} + \sigma_a^y \,\sigma_{a+1}^{y} +  \De \, (\sigma_a^z \,\sigma_{a+1}^{z} + 1)  \Big\} \, - \, \f{h}{2} \sul{a=1}{L} \sigma_a^z  \; .
\label{ecriture hamiltonien XXZ}
\enq
Here  $J>0$ stands for the exchange interaction, $\De \in {\mathbb R}$
is the anisotropy parameter, $h>0$ corresponds to the external magnetic field.
$\op{H}$ acts on the Hilbert space $\mf{h}_{XXZ}=\bigotimes_{a=1}^{L}\mf{h}_a$ with $\mf{h}_a \simeq \Cx^2$;
$\sg^{\a}$, $\a\in \{x, y, z \}$, are the Pauli matrices, and the operator
$\sg_{a}^{\a}$ acts as the Pauli matrix $\sg^{\a}$ on $\mf{h}_a$ and as the
identity on all the other spaces:
\beq
\sg_{a}^{\a} \, = \, \underbrace{ \e{id} \otimes \cdots \otimes \e{id} }_{a-1} \otimes \, \sg^{\a} \otimes \underbrace{ \e{id} \otimes \cdots \otimes \e{id} }_{L-a}  \;.
\enq
This model is equivalent to a system of free fermions at $\De=0$. It  also becomes trivial when $\De \tend +\infty$. However, for all other values of $\De$, it is
genuinely interacting.

\vspace{2mm}
The zero temperature dynamical two-point functions of this model are defined by
\beq
\big< \sg_1^{\a} \sg_{m+1}^{\a^{\prime}}(t)\big>_{0} \; = \; \lim_{L\tend +\infty} \Big( \Psi_{\e{g}}, \sg_1^{\a}\cdot  \sg_{m+1}^{\a^{\prime}}(t) \Psi_{\e{g}} \Big) \;,
 \quad \e{where} \quad \sg_{k}^{\a^{\prime}}(t) \; = \; \ex{ \i \op{H} t } \cdot \sg_{k}^{\a^{\prime}} \cdot \ex{ - \i \op{H} t }
\enq
while $\Psi_{\e{g}}$ stands for the ground state of $\op{H}$.  In their turn, the finite temperature correlators take the form
\beq
\big< \sg_1^{\a} \sg_{m+1}^{\a^{\prime}}(t)\big>_{T} \; = \; \lim_{L\tend +\infty}
                \Bigg\{ \f{ \e{tr}\Big[ \ex{- \f{1}{T}\op{H}} \sg_1^{\a}\cdot  \sg_{m+1}^{\a^{\prime}}(t)  \Big] }{ \mc{Z}_{L}  }   \Bigg\}
\qquad \e{with} \qquad   \mc{Z}_{L} \; = \; \e{tr}\Big[ \ex{- \f{1}{T}\op{H}}  \Big] \;.
\enq

The first explicit characterisation of thermal dynamical correlation functions has been achieved at the free fermion point $\De=0$. The first closed expression
was obtained in \cite{NiemejerEXactThermalCalculationLongDyn2PtFctXX} for the longitudinal case $\big< \sg_1^{z} \sg_{m+1}^{z}(t)\big>_{T}$.
The transverse case $\big< \sg_1^{\pm} \sg_{m+1}^{\mp}(t)\big>_{T}$ turned out to be much more involved, even at $\De=0$,
and a first, quite intricate representation, has been obtained in \cite{AbrahamBarouchMcCoy4ptcomputationtoGetlongTimeforTwo}.
Later a simpler representation for $\big< \sg_1^{\pm} \sg_{m+1}^{\mp}(t)\big>_{ 0}$
was found in \cite{MullerShrockDynamicCorrFnctsTIandXXAsymptTimeAndFourier} in terms of Painlev\'{e} transcendents.
A important progress was made in \cite{ColomoIzerginKorepinTognettiXX2ptFctsatGeneralTimeAndSpaceAndTempe,ColomoIzerginKorepinTognettiTempCorrFctXX}
where closed Fredholm determinant based representations were obtained for  $\big< \sg_1^{\pm} \sg_{m+1}^{\mp}(t)\big>_{T}$.
The integral kernels involved in the latter were of integrable type \cite{ItsIzerginKorepinSlavnovDifferentialeqnsforCorrelationfunctions,
SakhnovichFirstStepsOfElementsOfIntegrableIntOps}, what allowed the study of the long-time large-distance asymptotics of those correlation functions by
means of Riemann--Hilbert problem techniques \cite{ItsIzerginKorepinSlavnovTempCorrFctSpinsXY,JieLargetABofDynCorrFctsinXX}.

The case of dynamical correlators in a genuinely interacting chain, \textit{viz.} for $\De \not\in \{ 0, +\infty \}$, could only be achieved much later.
First representations were obtained for the zero-temperature case, which is structurally simpler. They were obtained for the so-called
massive regime, \textit{viz}. when $\De>1$, and followed from the computation of form factors of local operators within
the vertex operator approach \cite{JimboMiwaFormFactorsInMassiveXXZ}.
The case of the massless regime was dealt with a decade later, within the quantum inverse scattering formalism.  In the work
\cite{KitanineMailletSlavnovTerrasDynamicalCorrelationFunctions} a series representation for the zero temperature longitudinal spin-spin
correlation functions $\big< \sg_1^{\pm} \sg_{m+1}^{\mp}(t)\big>_0$  was obtained. The extension to the case of finite temperature was made possible thanks to
the quantum transfer matrix formalism,
introduced in full generality in \cite{KlumperNLIEfromQTMDescrThermoRSOSOneUnknownFcton}.
The work \cite{SakaiDynamicalAndTimeDependentCorrelatorsXXZ} proposed, for the first time, a closed expression for the
dynamical longitudinal spin-spin correlators of the XXZ spin-$1/2$ chain at finite temperature $\big< \sg_1^{z} \sg_{m+1}^{z}(t)\big>_{T}$.
More recently, an important development in the description of finite temperature dynamical correlators in the XXZ chain
took place within the thermal form factor approach, first developed for static correlators in
\cite{KozDugaveGohmannThermaxFormFactorsXXZ,KozDugaveGohmannThermaxFormFactorsXXZOffTransverseFunctions}.
Indeed, the work \cite{KozGohmannKarbachSuzukiFiniteTDynamicalCorrFcts} proposed a closed and fully explicit expression
for thermal dynamical two-point functions in integrable models related with a fundamental $R$-matrix,
the XXZ spin-$1/2$ chain being a paradigmatic example thereof. Recently, the approach was generalised to deal
with two-point functions of arbitrary multi-site operators in \cite{KozGohmannMinin2ptthermadynamicalMultiSiteCorrFcts}.
The representation obtained in \cite{KozGohmannKarbachSuzukiFiniteTDynamicalCorrFcts} showed its efficiency when closed and fully explicit expressions
for the zero temperature dynamical two-point functions in the massive regime, \textit{viz.}  $\De>1$,
were obtained in \cite{KozBabenkoGohmannSirkerSuzukiMassiceXXZFFExplicitDensityFullyComputedAndNumerics,KozBabenkoGohmannSuzukiMassiceXXZFFExplicitDensity}.
This was in stringent contrast
with the previously obtained formulae
\cite{KozDugaveGohmannSuzukiLargeVolumeBehaviourFFMassiveXXZ,KozDugaveGohmannSuzukiLowTFFSeriesMassiveXXZQTMandXXX,JimboMiwaFormFactorsInMassiveXXZ}.
The mentioned progress allowed, in particular, to obtain a particularly simple representation for the
spin conductivity \cite{KozGohmannSirkerSuzukiSpinConductivityXXZMassiveDSF} in that regime.

It is important to stress at this point that obtaining closed exact representations for the thermal dynamical correlation functions  is just a part of the full story. Indeed,
one should be able to extract from these all the physcially pertinent information one needs. Developing this second part of the story, at least
to some extent, is the purpose of this work. More precisely, we focus on the extraction -directly from the series- of the large-distance long-time asymptotic behaviour
of transverse two-point functions. As mentioned, this is of special interest to universality features of the chain, at least in what concerns the low-temperature
regime. There, dynamical correlators are expected to exhibit a universal structure in their leading long-time and large-distance asymptotic behaviour,
captured by the thermal non-linear Luttinger model \cite{KarraschPereiraSirkerLowTDynamicsofNLLLAndApps}. Hence, one of the motivations of this work is to pave the way
for testing these predictions from a lattice calculation. Of course, a particular case of the asymptotic regime corresponds to the long-distance regime for the static correlation functions
in which case the universal structure is captured by the Luttinger-Liquid model. This has been established from an \textit{ab initio} study of the model's correlation functions
for the non-linear Schrödinger model and the XXZ chain in previous works, respectively
\cite{KozMailletSlaLongDistanceTemperatureNLSE,KozMailletSlaLowTLimitNLSE} and \cite{KozDugaveGohmannThermaxFormFactorsXXZ,KozDugaveGohmannThermaxFormFactorsXXZOffTransverseFunctions}.
However, the case of non-vanishing time is totally open to the best of our knowledge.

The series of multiple-integral representations  obtained in \cite{KozGohmannKarbachSuzukiFiniteTDynamicalCorrFcts} also demonstrated its
efficiency for dealing with the  long-time and large-distance asymptotics at finite temperature, in the space-like regime,
at the free fermion point $\De=0$, see \cite{KozGohmannSuzukiXXTransverseSpaceLikeAsymptotics}. While the works
\cite{ItsIzerginKorepinSlavnovTempCorrFctSpinsXY,JieLargetABofDynCorrFctsinXX} already considered the calculation of these asymptotics,
the main progress resided in the huge simplification in its derivation. Indeed,
the works \cite{ItsIzerginKorepinSlavnovTempCorrFctSpinsXY,JieLargetABofDynCorrFctsinXX} relied on Riemann--Hilbert problem techniques
which demanded quite a few steps in the analysis. In its turn, the analysis of \cite{KozGohmannSuzukiXXTransverseSpaceLikeAsymptotics}
was based on very simple contour deformations and residue calculations. This suggested that the series at generic $\De\in \intoo{-1}{1}$
would enjoy the very same properties. This is indeed the main conclusion of the present work. \textit{The leading space-like asymptotics
at low-temperature of dynamical two-point functions can be obtained from simple contour deformations}.
In fact, we find that for the transverse two-point functions, the leading space-like asymptotics are obtained
from the Eigenvalue of the quantum transfer matrix that is closest to the dominant one.

Our main result takes the following form.

\begin{conj}
\label{Conjecture pple article}

There exists a constant $\mf{c}>0$, possibly depending on $T$, such that, when $m \tend +\infty$ and $|t| \, <\, \mf{c}\,  m$,
the transverse thermal dynamical two-point function admits the asymptotic behaviour
\beq
\big< \sg_1^- \sg_{m+1}^+(t) \big>_{T} \, = \, (-1)^m \msc{B}^{-+}(T) \ex{\i m  \De_{\e{dom}}} \Big(  1 \,  +\, \e{o}(1) \Big)
\enq
Here, $\msc{B}^{-+}(T) = \tf{ T  \msc{A}^{-+}_{\e{dom}}  }{ u^{\prime}_{\e{dom}}( x ) }$.
The amplitude $ \msc{A}^{-+}_{\e{dom}} $ is a functional of $u_{\e{dom}}$ and depends on $x$.
It is given in terms of a product  of two
thermal form factors, one associated with the $\sg_1^{-}$ operator and the other with $\sg_1^+$. Further,
the parameter $x$ and the function $u_{\e{dom}}$ solve the coupled systems of equations
\beqa
 u_{\e{dom}}(x) & = & -\i\pi T \;,  \Re(x) >0 \;, \label{equation position dominante trou} \\
u_{\e{dom}}(\la  ) & = & \veps_{0}(\la) \, - \,  \i \pi  T
\, + \, \i T \th(\la-x) \, - \, T \Int{ \msc{C}_{ u_{\e{dom}} } }{} \hspace{-2mm}\dd \mu\,  K(\la-\mu) \msc{L}n\big[1+\ex{-\frac{1}{T}u_{\e{dom}} } \big](\mu   ) \;.
\label{ecriture NLIE pour solution sous dominante}
\eeqa
There $\msc{C}_{u}$ is an integration contour defined in terms of $u$, and hence whose construction is part of the non-linear problem.
Its description may be found in the core of the paper, Section \ref{SousSection NLIE}.  Likewise, $\msc{L}n$ is a determination of the logarithm
defined later on, in \eqref{definition logarithm de 1+a}. Finally, $\veps_0$ is the bare energy defined in \eqref{definition bare energy}.

In order to state the formula for the correlation length, we still need to introduce another solution to a non-linear integral equation,
the one describing the dominant Eigenstate of the quantum transfer matrix:
\beq
     u_{\e{D}}\big( \xi  \big) \, = \,  \veps_0(\xi)
     \; - \; T\Oint{  \msc{C}_{u_{\e{D}}}  }{} \dd \la \: K(\xi-\la)
        \msc{L}\mathrm{n}_{ \msc{C}_{ u_{\e{D}} } }
        \Big[ 1+  \ex{ - \f{1}{T}u_{\e{D}}  } \, \Big](\la) \;.
\enq
The above being settled, the inverse correlation length takes the form $ \De_{\e{dom}} \, = \, \mc{P}_{\e{dom}} \, + \, \f{t}{m} \mc{E}_{\e{dom}}$ in which
\beqa
\mc{P}_{\e{dom}} & = &    - \,     p_0(x)
\, - \,  \Oint{ \msc{C}_{ u_{\e{dom}} } }{ } \f{ \dd \la }{ 2 \i \pi } \:
	p_0^{\prime}(\la ) \,
  \msc{L}\mathrm{n}_{ \msc{C}_{ u_{\e{dom}} } } \Big[ 1+  \ex{ - \f{1}{T}u_{\e{dom}} } \, \Big](\la)
                \, + \,  \Oint{ \msc{C}_{ u_{\e{D}} } }{ } \f{ \dd \la }{ 2 \i \pi } \:
	p_0^{\prime}(\la ) \, \msc{L}\mathrm{n}_{ \msc{C}_{ u_{\e{D}} } }\Big[ 1+  \ex{ - \f{1}{T}u_{\e{D}}   } \, \Big](\la) \; ,   \nonumber \\
 \mc{E}_{\e{dom}} &  = &   - \,    \veps_0(x)
\, - \,  \Oint{ \msc{C}_{ u_{\e{dom}} } }{ } \f{ \dd \la }{ 2 \i \pi } \:
	\veps_0^{\prime}(\la ) \,
  \msc{L}\mathrm{n}_{ \msc{C}_{ u_{\e{dom}} } } \Big[ 1+  \ex{ - \f{1}{T}u_{\e{dom}} } \, \Big](\la)
         \, + \,      \Oint{ \msc{C}_{  u_{\e{D}}  } }{ } \f{ \dd \la }{ 2 \i \pi } \:
	\veps_0^{\prime}(\la ) \,
  \msc{L}\mathrm{n}_{ \msc{C}_{ u_{\e{D}} } }\Big[ 1+  \ex{ - \f{1}{T}u_{\e{D}}   } \, \Big](\la)         \;.
\nonumber
\eeqa
Above, $p_0$ refers to the bare momentum defined in \eqref{definition dressed momentum}.
In the low-$T$ regime, one has moreover that
\beq
\De_{\e{dom}} \; = \; \f{\i\pi T }{  2 \op{v}_F Z^2(q) } \Big( 1+\e{O}(T)\Big) \; ,
\label{ecriture valued inverse correlation length at low T}
\enq
in which $\op{v}_F$ refers to the Fermi velocity \eqref{definition Fermi velocity}, $Z$ is the dressed charge \eqref{definition dressed charge} and $q$ stands for the endpoint of the Fermi zone, \textit{c.f.}
below of \eqref{definition energie habille et energie nue}.

\end{conj}

It appears appropriate, at this stage, to comment on the very form, provided by the conjecture, of the long-distance and large-time asymptotics
of the transverse function in a cone of the space-like regime.
First of all, the asymptotics are expected to hold throughout the whole range of temperatures $T \in \intoo{0}{+ \infty}$, provided that $T$
is fixed -or at least bounded from below away from $0$-. In the particular case of $t=0$, which can be directly set in the formulae, one simply recovers that
the leading asymptotics are given by the infinite Trotter number limit of the ratio of the first sub-dominant to the dominant Eigenvalue of the quantum transfer matrix.
This has been established in \cite{KozDugaveGohmannThermaxFormFactorsXXZ,KozDugaveGohmannThermaxFormFactorsXXZOffTransverseFunctions}
and argued -prior to that research- in the works \cite{SuzukiInoueMoreDvpmtInterchangeabilityTrotterAndVolumeLimitInPartFcton,TakahashiThermoXXZInfiniteNbrRootsFromQTM}-. What the conjecture stresses is that, up to certain modifications,
structurally the answer is similar, at least when the time $|t|$ is not too large in respect to the distance $m$, this structure is preserved in that
the leading asymptotics are directly inferred from the spectral data relative to the Trotter limit of the
dominant and first subdominant Eigenvalue of the quantum transfer matrix. Of course, when the time becomes larger and larger in magnitude,
other effects may come into play - in particular saddle points appearing at certain critical values of the excitations' dispersion relations-
which will modify the behaviour. In particular, taken the form of the answer obtained by Riemann--Hilbert techniques for the free fermionic XX chain \cite{ColomoIzerginKorepinTognettiTempCorrFctXX},
such saddle-point will generate a power-law behaviour in the distance when $|\tf{t}{m}|$ is large enough.

One may also wonder how the conjecture relates to the universal
behaviour expected to arise in the $T\tend 0^+$ regime when $mT, tT$ converges to some finite, possibly zero, values.
As inferred from \eqref{ecriture valued inverse correlation length at low T}, the dominant inverse correlation length converges to $0$ as $T\tend 0^+$
so that $m \De_{\e{dom}}$ approaches a \textit{finite} values in the scaling regime.
Worse, the amplitude will approach zero algebraically in $T$, see \cite{KozDugaveGohmannThermaxFormFactorsXXZ,KozDugaveGohmannThermaxFormFactorsXXZOffTransverseFunctions}.
This means that the sole leading term of the asymptotics \textit{does not} capture any more the scaling behaviour: the subleading terms will
all be of the same order of magnitude and their joint resummation will produce a counter term to the vanishing of the amplitude so that, all-in-all,
once that the resummation is performed, one obtains a finite behaviour in the scaling regime. This was the mechanism that arose in the Luttinger liquid
scaling regime corresponding to static correlators that was obtained in
\cite{KozMailletSlaLongDistanceTemperatureNLSE,KozMailletSlaLowTLimitNLSE,KozDugaveGohmannThermaxFormFactorsXXZ,KozDugaveGohmannThermaxFormFactorsXXZOffTransverseFunctions}.
The subtleties of the resummation arising in the genuinely dynamical case will, however, surely go beyond the framework developed in
the mentioned works due to the presence of saddle-point contributions.

The paper is organised as follows. Section \ref{Section Thermal FF series} introduces all the technical ingredients that are necessary for setting up the
thermal form factor series expansion for the dynamical correlation functions of the XXZ spin-$\tf{1}{2}$ chain. In Subsection \ref{SousSection LIE}
we introduce the solutions to linear integral equations. Next, in Subsection \ref{SousSection NLIE}, we discuss the non-linear integral
equations which describe the spectral properties of the quantum transfer matrix in the infinite Trotter number limit.
Finally, in Subsection \ref{SousSection dynamical thermal FF series}, we discuss the thermal form factor expansion \textit{per se}.
Section \ref{Section Large m t analysis} is devoted to the analysis of the leading behaviour of the transverse thermal dynamical
two-point functions starting from their thermal form factor series expansion. In Subsection \ref{SousSection Reecriture de la series themal FF},
we recast the thermal form factor series in a form that is adapted for its large-$m$ long-$t$ analysis. The latter is discussed in Subsection
\ref{SousSection large m t analysis}. Then, in Subsection \ref{SousSection Free Fermion Results}, we recall our previous result
pertaining to the  large-$m$ long-$t$ analysis of the thermal form factor series at the free fermion point and recast it in a form where
an identification of constants in terms of thermal form factors is apparent. Finally, in Subsection \ref{SousSection Conjecture for cone in space like regime},
we present our conjecture on the leading term in the large-$m$ and long-$t$, $|t|\leq \mf{c} \, |m|$, asymptotic behaviour of dynamical
thermal two-point functions in the XXZ chain.

\section{The thermal form factor series for dynamical two-point functions}
\label{Section Thermal FF series}

\subsection{Solutions to linear integral equations}
\label{SousSection LIE}

It is well established nowadays that numerous properties of the XXZ chain in an external magnetic field $h>0$
are described in terms of solutions to linear integral equations of the type
\beq
     \Big(\e{id} + \op{K} \Big)[f](\la) \; =
        \;f(\la) \; + \; \Int{-q}{q}  \dd \mu \: K(\la-\mu) f(\mu)   = g(\la)\;.
\label{definition op int id + K}
\enq
In the Bethe Ansatz formalism, $g$ is called the bare quantity and the solution $f$ the dressed one.
The operator $\e{id} + \op{K}$ is understood to act on $L^2\big( \intff{-q}{q} \big)$ and has an explicit integral kernel given by
\beq
K(\xi) \, = \, \f{ \sin(2\zeta)  }{ 2 \pi \sinh(\xi-\i\zeta) \sinh(\xi+\i\zeta)  }\;.
\label{definition K}
\enq
The parameter $q$ depends on the value of the magnetic field and is such that the Fermi zone of the model adapts itself to the magnitude of the magnetic field.
To define it more specifically, one first needs to introduce the bare energy
\beq
\veps_{0}(\la)  \, = \,  h -  \f{ 2 J \sin^2(\zeta) }{ \sinh\big(\la + \frac{\i}{2}\zeta \big)  \sinh\big(\la - \frac{\i}{2}\zeta \big)  } \; .
\label{definition bare energy}
\enq
One can show that the operator $\e{id} + \op{K} $ is actually invertible on $L^2\big( \intff{-Q}{Q} \big)$ for any $0 \leq Q \leq +\infty$,
see \cite{KozDugaveGohmannThermoFunctionsZeroTXXZMassless,KozProofOfDensityOfBetheRoots,YangYangXXZStructureofGS}.
This then gives rise to the family of special functions $\veps(\la\mid Q)$ defined as solutions to the below linear
integral equation
\beq
     \veps(\la\mid Q) \,
        + \, \Int{-Q}{Q} \dd \mu \: K\big(\la-\mu  \big) \, \veps(\mu\mid Q)
	  \; = \;  \veps_0(\la)  \;.
\label{definition energie habille et energie nue}
\enq
Here, $ \veps_0$ is the bare energy introduced in \eqref{definition bare energy}.
For $0 < h < 4J (1 + \De)$, the endpoint of the Fermi zone $q$ is defined as
the unique \cite{KozDugaveGohmannThermoFunctionsZeroTXXZMassless,KozFaulmannGohmannDressedEnergyInComplexPlane} positive solution $q$
to the equation $\veps(Q\mid Q)=0$. The associated solution of
\eqref{definition energie habille et energie nue} is called the dressed
energy and is denoted $\veps(\la) \equiv \veps(\la \mid q )$.

The dressed energy enjoys several properties which play an important role in the
analysis of the low-temperature behaviour of the spectrum of the quantum transfer matrix \cite{KozFaulmanGohmannLowTNLIERigourousAnalysisForQTMMasslessRegime}.
In particular, it is shown in \cite{KozFaulmanGohmannLowTNLIERigourousAnalysisForQTMMasslessRegime}
that, for $0 < \zeta < \pi/2$, the map $\veps: \msc{U}_{\veps} \setminus \big\{ 0, \i\tfrac{\pi}{2} \big\}
\tend \veps\big( \msc{U}_{\veps} \setminus \big\{ 0, \i\tfrac{\pi}{2} \big\} \big)$ with
\beq
     \msc{U}_{\veps} \; = \;
        \Big\{ z \in \Cx \; : \; - \tfrac{\pi}{2} < \Im(z) \leq \tfrac{\pi}{2} \Big\}
	\setminus \bigcup_{\ups= \pm} \Big\{ \intff{-q}{q} + \i \ups \zeta \Big\} \; ,
\label{definition ensemble de support pour dble rec}
\enq
is a double cover. Moreover, it follows from the analysis carried out in
\cite{KozFaulmanGohmannLowTNLIERigourousAnalysisForQTMMasslessRegime,KozFaulmannGohmannDressedEnergyInComplexPlane} that $\veps$ is invertible
in a neighbourhood of the curves $\{ \pm \Re(\la) >0, \Re[\veps(\la)]=0 \}$.

In order to discuss our result, we will need two more special functions
solving linear integral equations driven by  $\e{id} + \op{K}$. First of all, we define
 the dressed charge $Z$ as the solution to
\beq
     Z(\la)\, + \,  \Int{-q}{q}  \dd \mu \: K(\la-\mu)\,  Z(\mu ) \, = \,  1 \;.
\label{definition dressed charge}
\enq
In its turn, the the dressed momentum $p$ corresponds to the unique solution to
\beq
     p(\la)\, + \, \Int{-q}{q} \frac{\dd \mu}{2 \pi} \: \theta (\la-\mu)\,  p^{\prime}(\mu )
        \, = \, p_0(\la) \;  \qquad \e{with} \qquad \quad  p_0(\la) \, = \, \i \ln \bigg(  \f{ \sinh(\i\tf{\zeta}{2}+ \la) }{  \sinh(\i\tf{\zeta}{2} - \la) } \bigg) \;.
\label{definition dressed momentum}
\enq
The above equation contains the bare phase defined as
\beq
\label{definition_theta}
     \th(\la) \; = \; \left\{
        \ba{ccc}
        \i \ln \bigg( \f{ \sinh(\i\zeta+\la) }{ \sinh(\i\zeta-\la) }\bigg)
	& \text{for} & |\Im(\la)| \, < \, \e{min}(\zeta, \pi-\zeta)  \vspace{3mm}\\
        -\pi \e{sgn}(\pi-2\zeta)   +\i \ln \bigg( \f{ \sinh(\i\zeta+\la) }{ \sinh(\la - \i\zeta) }\bigg)
        & \text{for} &  \e{min}(\zeta, \pi-\zeta) \, < \, |\Im(\la)| \, < \, \tf{\pi}{2}
        \ea \right. \;.
\enq
Above $"\ln"$ refers to the principal branch of the logarithm.
%
%
%
%
%
%
%
%

\subsection{The non-linear integral equation approach to the spectrum of the quantum transfer matrix}
\label{SousSection NLIE}

The quantum transfer matrix is an auxiliary tool allowing one to deal efficiently with the thermodynamics
of quantum integrable models. It is at the root of the representation of the
thermal dynamical two-point functions by means of  series of multiple integrals that we shall describe and analyse later on.
The infinite Trotter number limit of the spectrum of the quantum transfer matrix plays an
important role in the description of that series. We will thus review its characterisation
by means of non-linear integral equations.
We will moreover focus on the description of the low-$T$ regime since it is the one pertinent for the handlings to come.
We refer to \cite{KozFaulmanGohmannLowTNLIERigourousAnalysisForQTMMasslessRegime} for a thorough discussion.
Below, we shall discuss the whole construction in the regime $0 \, \leq\,  \De \, <\,  1$. The reasons are twofold. On the one hand, it is
in this range of parameters that the solvability theory of the non-linear integral equations of interest is amenable to a fully rigorous
analysis. On the other hand, in this regime, the limiting spectrum of the quantum transfer matrix admits a particularly simple description
in terms of particle-hole excitations. We stress that the regime $-1 < \De < 0$ may also be dealt with, although not yet in full rigour.
We refer to \cite{KozFaulmanGohmannLowTNLIERigourousAnalysisForQTMMasslessRegime} for more details.
Also, when $-1 < \De < 0$, one observes string solutions whose appropriate dealing with would unnecessarily
overburden the main ideas of the analysis that we develop in this work.

We shall start by describing the class of functions $u$ over which one solves the non-linear integral
equation. This will then allow us to associate with such $u$ a contour $\msc{C}_{u}$. This contour then arises in the very formulation
of the non-linear integral equation as well as in all integral representations for the physical observables.

 We focus on solutions $u$ that are holomorphic in some open neighbourhood  $\mc{V}_{\pm q}$ of the Fermi points $\pm q$
 which contains the open discs $\op{D}_{\pm q , \eps}$ of $T$-independent radius $\eps>0$ and such that
\beq
u: \mc{V}_{\pm q} \tend \op{D}_{0,\varrho}\, , \quad  \e{with} \quad  \varrho>0
\enq
and $T$ independent, is a biholomorphism. In particular, such $u$ admit a unique zero $q^{(+)}_{u }$, resp. $q^{(-)}_{u }$, \textit{viz}. $u\big(  q^{(\pm)}_{ u }   \big)
\, = \, 0 $, located in $\mc{V}_{q}$, resp. $\mc{V}_{-q}$.  These zeroes are such that $\pm \Re\big[ u^{\prime}\big(  q^{(\pm)}_{ u }  \big) \big]
> c> 0$ for some $c>0$, uniformly in $T$ small enough. Finally, $\la \mapsto 1-\ex{-\f{1}{T}u(\la) }$ is piecewise meromorphic on $\Cx$
and $\i\pi$-periodic.

Given $u$ as above, we define a contour $\msc{C}_{u}$ satisfying the requirements
\begin{itemize}
\item[{\rm i)}] $\msc{C}_{ u }$ passes through two zeroes $q^{(\pm)}_{u }$ of
$1-\ex{-\frac{1}{T}u}$  satisfying $u\big(  q^{(\pm)}_{ u }   \big)
\, = \, 0 $ and these are the only zeroes of $1-\ex{-\frac{1}{T}u}$ on  $\msc{C}_{ u }$;

\item[\rm{ii)}] there exists $J_{\de}^{(\pm)}\subset \msc{C}_{ u }$ such that
$u\big( J_{\de}^{(-)}   \big)= \intff{-\de}{\de} $ and
$u\big(J_{\de}^{(+)} \big)= \intff{\de}{-\de} $,
for some $\de>0$, possibly depending on $T$, but such that $\tf{\de}{T} > -C \ln T$ as $T \tend 0^+$, for some $C>0$;

\item[\rm{iii)}]  the complementary set $J_{\de} \, = \, \msc{C}_{ u }
\setminus \Big\{ J_{\de}^{(-)} \cup J_{\de}^{(+)}  \Big\}$ is such that
$\big| \Re\big[ u(\la ) \big] \big|  > \tf{\de}{2}$  for all $\la \in J_{\de}$.

\end{itemize}
Above, we have denoted by $\intff{a}{b}$ the oriented segment run through from $a$ to $b$.
As demonstrated in \cite{KozFaulmanGohmannLowTNLIERigourousAnalysisForQTMMasslessRegime}, the above conditions
imply that $1+\ex{-\f{1}{T}u}$ has zero index with respect to $\msc{C}_u$:
\beq
 \Oint{ \msc{C}_u }{ } \f{ \dd \mu }{ T } \:
        \f{ -  u^{\, \prime}\!(\mu) }{ 1+  \ex{ \f{1}{T}  u \, ( \mu )} } \; = \; 0 \;.
\label{zero monodromy condition}
\enq

We are now in a position to introduce the non-linear integral equation which describes the largest Eigenvalue of the quantum transfer matrix
in the infinite Trotter number limit:
\beq
     u_{\e{D}}\big( \xi  \big) \, = \,  \veps_0(\xi)
     \; - \; T\Oint{  \msc{C}_{u_{\e{D}}}  }{} \dd \la \: K(\xi-\la)
        \msc{L}\mathrm{n}_{ \msc{C}_{ u_{\e{D}} } }
        \Big[ 1+  \ex{ - \f{1}{T}u_{\e{D}}  } \, \Big](\la) \;.
\label{ecriture eqn NLI pour dom state}
\enq
Here, $\veps_0$ is the bare energy, and the integral equation holds for  $\xi$ belonging to $\msc{C}_{ u_{\e{D}} }$.
Finally, for any function $u$ and given $\nu \in \msc{C}_u$, we agree to define the logarithm as
\beq
     \msc{L}\mathrm{n}_{  \msc{C}_u }
        \Big[ 1+  \ex{ - \f{1}{T}u  } \, \Big](\nu) \, = \,
        \Int{\varkappa}{v} \f{ \dd \mu }{ T } \:
        \f{ -  u^{\, \prime}\!(\mu) }{ 1+  \ex{ \f{1}{T}  u \, ( \mu )} }
     \; + \; \ln \Big[ 1+  \ex{- \f{1}{T} u ( \varkappa )} \, \Big] \;.
\label{definition logarithm de 1+a}
\enq
Above, $\varkappa$ is some point on $\msc{C}_u$.
Here we stress that, owing to the zero monodromy condition, the  definition of the logarithm does not depend on $\varkappa$.

This solution provides a simple integral representation of the dominant Eigenvalue in the infinite Trotter number limit:
\beq
\La_{\e{D}}
        \, = \,  \ex{  \f{h}{2T}   -\f{2J}{T}\cos(\zeta) }
	\cdot  \ex{ \i \mc{P}_{\e{D}} } \qquad \e{where} \qquad
\mc{P}_{\e{D}} \, = \, - \Oint{ \msc{C}_{ u_{\e{D}} } }{ } \f{ \dd \la }{ 2 \i \pi } \:
	p_0^{\prime}(\la ) \,
     \msc{L}\mathrm{n}_{ \msc{C}_{u_{\e{D}}} }
        \Big[ 1+  \ex{ - \f{1}{T}u_{\e{D}} } \, \Big](\la)  \;.
\enq

The description of subdominant Eigenvalues demands to deal with more complex non-linear integral equations
containing an extra driving term
\beq
     \Theta (\xi \, |\, \bs{x}_{n} , \bs{y}_{m} ) \, = \, \sul{ a=1 }{ m } \th (\xi- y_a)
\, - \, \sul{ a=1 }{ n } \th (\xi- x_a)
\enq
which depends on two sets of auxiliary parameters
\beq
\bs{x}_n \; = \; \big( x_1,\dots, x_n \big)^{\op{t}}  \qquad \e{and} \qquad  \bs{y}_m \; = \; \big(y_1,\dots, y_m  \big)^{\op{t}} \;.
\enq
The driving term is expressed with the help of the bare phase introduced in \eqref{definition_theta}.

The non-linear integral equation of interest takes the form
\beq
     u\big( \xi \mid  \bs{x}_{n} , \bs{y}_{m} \big) \, = \,  \veps_0(\xi)  \; - \; \i \pi \mf{s}  T
        \; - \; \i T \Theta (\xi \, | \,  \bs{x}_{n} , \bs{y}_{m})
     \; - \; T\Oint{  \msc{C}_u  }{} \dd \la \: K(\xi-\la)
        \msc{L}\mathrm{n}_{ \msc{C}_u }
        \Big[ 1+  \ex{ - \f{1}{T}u (* \mid \, \bs{x}_{n} , \bs{y}_{m}  ) } \, \Big](\la) \;.
\label{ecriture eqn NLI forme primordiale}
\enq
The integral equation holds for  $\xi$ belonging to $\msc{C}_u$ and the logarithm is defined as in \eqref{definition logarithm de 1+a}.
The integer $\mf{s} \, = \,  n-m\in \mathbb{Z}$ corresponds to the pseudo-spin of the associated excited  state.
This equation admits a unique solution for a large range of points $\bs{x}_n, \bs{y}_m$. We refer to   \cite{KozFaulmanGohmannLowTNLIERigourousAnalysisForQTMMasslessRegime}
for more details.

\vspace{2mm}

Now, in order to access the spectrum of the quantum transfer matrix in the infinite Trotter number limit, one starts from a solution $u$ to
\eqref{ecriture eqn NLI forme primordiale} and imposes quantisation conditions on the parameters  $\bs{x}_n, \bs{y}_m$.
These should correspond to simple roots of the equations
\beq
\ex{- \f{1}{T}u( \mf{x}_a \mid  \bs{\mf{x}}_{n} , \bs{\mf{y}}_{m} )} \; = \; - 1 \;, \quad a=1,\dots, n \quad \e{and} \quad
\ex{- \f{1}{T}u( \mf{y}_a \mid  \bs{\mf{x}}_{n} , \bs{\mf{y}}_{m} )} \; = \; - 1 \;, \quad a=1,\dots, m \;,
\enq
with $\mf{x}_a$ belonging to the interior of $\msc{C}_u$ and $\mf{y}_a$ to its exterior, modulo $\i\pi$.
Above, $\bs{\mf{x}}_n$, resp. $\bs{\mf{y}}_m$, is the vector with coordinates $\mf{x}_a$, resp. $\mf{y}_a$. Moreover,
in order to obtain a \textit{per se} Eigenvalue, it should also hold that
\beq
\big(\Dp{\la} u\big)( \mf{x}_a \mid  \bs{\mf{x}}_{n} , \bs{\mf{y}}_{m} )\not=0 \; , \quad a=1,\dots, n \quad \e{and} \quad
\big(\Dp{\la} u\big)( \mf{y}_a \mid  \bs{\mf{x}}_{n} , \bs{\mf{y}}_{m} )\not=0 \;, \quad a=1,\dots, m \;.
\enq

Then, given a solution to the above problem, the associated infinite Trotter-number limit of the Eigenvalue of the quantum transfer matrix
takes the explicit form
\beq
     \La\big(\bs{\mf{x}}_{n} , \bs{\mf{y}}_{m} \big)
        \, = \, (-1)^{\mf{s}}  \cdot \ex{  \f{h}{2T}   -\f{2J}{T}\cos(\zeta) }
	\cdot  \ex{ \i \mc{P} ( \bs{\mf{x}}_{n} , \bs{\mf{y}}_{m}   ) } \cdot  \ex{ \i \mc{P}_{\e{D}} } \;.
\label{ecriture forme explicite basse T pour  vp QTM}
\enq
Here $\mc{P}\big( \bs{\mf{x}}_{n} , \bs{\mf{y}}_{m}  \big)$ corresponds to the effective momentum carried by the Eigenvalue $\La$.
It is expressed in terms of $u$, for any parameters $\bs{x}_n, \bs{y}_{m}$ as
\beq
\mc{P}\big( \bs{x}_{n} , \bs{y}_{m}  \big) \; = \; \sul{a=1}{m}  p_0(y_a) \, - \, \sul{a=1}{n}   p_0(x_a)
\, - \,  \Oint{ \msc{C}_{ u } }{ } \f{ \dd \la }{ 2 \i \pi } \:
	p_0^{\prime}(\la ) \,
     \msc{L}\mathrm{n}_{ \msc{C}_u }
        \Big[ 1+  \ex{ - \f{1}{T}u (* \mid \, \bs{x}_{n} , \bs{y}_{m}  ) } \, \Big](\la) \, - \, \mc{P}_{\e{D}} \;.
\label{definition impulsion totale}
\enq

Later on, we shall need another spectral observable related to the quantum transfer matrix, the so-called effective energy carried by
the Eigenvalue $\La$. It is obtained as a certain infinite Trotter-number limit involving the finite Trotter counterpart of the
Eigenvalues $\wh{\La}$ and $\wh{\La}_{\e{D}}$. It is defined as
\beq
\mc{E}\big( \bs{x}_{n} , \bs{y}_{m}  \big) \; = \; \sul{a=1}{m}  \veps_0(y_a) \, - \, \sul{a=1}{n}   \veps_0(x_a)
\, - \,  \Oint{ \msc{C}_{ u } }{ } \f{ \dd \la }{ 2 \i \pi } \:
	\veps_0^{\prime}(\la ) \,
     \msc{L}\mathrm{n}_{ \msc{C}_u }
        \Big[ 1+  \ex{ - \f{1}{T}u (* \mid \, \bs{x}_{n} , \bs{y}_{m}  ) } \, \Big](\la) \, - \, \mc{E}_{\e{D}} \;,
\label{definition energie totale}
\enq
where
\beq
\mc{E}_{\e{D}} \, = \, \, - \,  \Oint{ \msc{C}_{ u_{\e{D}} } }{ } \f{ \dd \la }{ 2 \i \pi } \:
	\veps_0^{\prime}(\la ) \,
     \msc{L}\mathrm{n}_{ \msc{C}_{u_{\e{D}}} }
        \Big[ 1+  \ex{ - \f{1}{T}u_{\e{D}} } \, \Big](\la) \;.
\enq
It was established in \cite{KozFaulmanGohmannLowTNLIERigourousAnalysisForQTMMasslessRegime} that $\mc{P}\big( \bs{x}_{n} , \bs{y}_{m}  \big)$
and $\mc{E}\big( \bs{x}_{n} , \bs{y}_{m}  \big) $ admit the low-$T$ expansions
\beqa
\mc{P}\big( \bs{x}_{n} , \bs{y}_{m}  \big) & = & \sul{a=1}{m}  p(y_a) \, - \, \sul{a=1}{n}   p(x_a)  \; + \; \e{O}(T) \; ,  \label{ecriture DA P}\\
\mc{E}\big( \bs{x}_{n} , \bs{y}_{m}  \big) & = &   \sul{a=1}{m}  \veps(y_a) \, - \, \sul{a=1}{n}   \veps(x_a)  \; + \; \e{O}(T) \;  .
\label{ecriture DA E}
\eeqa
%
%
%



\subsection{Thermal form factor series expansion for transverse dynamical two-point functions}
 \label{SousSection dynamical thermal FF series}

We are now in position to introduce the thermal form factor series expansion for dynamical correlation functions which was first obtained in
\cite{KozGohmannKarbachSuzukiFiniteTDynamicalCorrFcts} and recently extended to multi-site operators in \cite{KozGohmannMinin2ptthermadynamicalMultiSiteCorrFcts}.
As already mentioned, we shall restrict our discussion to the $0 \leq \De <1$ regime, so as not to overburden the handlings, thus allowing
us to expose more transparently the core ideas behind our analysis. The main reason for this choice, \textit{c.f.}
\cite{KozFaulmanGohmannLowTNLIERigourousAnalysisForQTMMasslessRegime} for details, is that when $-1 < \De < 0$, on top of particle-hole excitations,
there also arise string solutions. This makes the description of the spectrum of the quantum matrix, and hence of the thermal form factor series, more technical.
In particular, the integration contour $\msc{C}_{n}$ that will be introduced in \eqref{definition C n Thermal FF series contour} will take a more involved form.
However, in the  end, the conclusions of the analysis will remain the same. We leave the corresponding details to the interested reader.

In order to describe this series, we first fix some integer $n \geq 1$ and introduce $n$, resp. $n-1$, dimensional vectors
\beq
\bs{x}_n \, = \, (x_1,\dots, x_n)^{ \op{t} }   \qquad \e{and} \qquad \bs{y}_{n-1} \, = \, \big(y_1,\dots, y_{n-1} \big)^{ \op{t} }
\enq
with coordinates lying uniformly away from $\pm \i \tf{\zeta}{2}$.
With this choice, we gather the two vectors into
\beq
\bs{w}=\big( \bs{x}_n, \bs{y}_{n-1} \big)^{\op{t}}
\label{definition vecteur parametres integres}
\enq
and consider $u(\la\mid \bs{w})$, the unique solution to the non-linear integral equation \eqref{ecriture eqn NLI forme primordiale} associated with the choice of parameters
$\bs{w}$. We denote by $q^{(\pm)}_{\bs{w}}$ the associated two solutions to $u\big( q^{(\pm)}_{\bs{w}} \mid \bs{w} \big)=0$ located in a $\e{O}(T)$
neighbourhood of $\pm q$. Observe that the curve
\beq
\msc{C}_{\bs{w}}\, = \,\Big\{ \la \; : \; \Re\big[  u(\la\mid \bs{w}) \big]=0 \quad \e{and} \quad |\Im(\la)| \leq \f{\pi}{2} \Big\}
\label{definition courbe C w}
\enq
is a small, $\e{O}(T)$ deformation of the curve
\beq
\msc{C}_{\veps}\, = \, \Big\{ \la \; : \; \Re\big[  \veps(\la) \big]=0 \Big\} \, .
\label{definition contour C veps}
\enq
As such, it consists of two branches:
\begin{itemize}
\item one $\msc{C}^{(-)}_{  \bs{w} }$ starting from $\i\tf{\zeta}{2}$ passing in an $\e{O}(T)$ vicinity of $-q$ through the point $q^{(-)}_{  \bs{w} }$
and then joining $-\i\tf{\zeta}{2}$. Along this curve $\Im\big[  u(\la\mid  \bs{w} ) \big]$
runs from $-\infty$ to $+\infty$.
\item One $\msc{C}^{(+)}_{  \bs{w} }$ starting from $-\i\tf{\zeta}{2}$ passing in an $\e{O}(T)$ vicinity of $q$ through
the point $q^{(+)}_{  \bs{w} }$ and then joining $\i\tf{\zeta}{2}$. Along this curve $\Im\big[  u(\la\mid  \bs{w} ) \big]$
runs from $-\infty$ to $+\infty$.

\end{itemize}
In the following, we denote by $u_{\pm}$ the restrictions of $u$ to an open neighbourhood of $\msc{C}^{(\pm)}_{ \bs{w} }$ in $\{\la \, : \, \pm   \Re[\la] > 0 \}$,
so that these become biholomorphisms.  Then, we define
\beq
\psi^{(h)}\big( \la \mid  \bs{w} \big) \; = \; \left\{ \ba{cc} u_+\big( \la \mid  \bs{w}  \big) \; ,  & \Im\big[ u \big( \la \mid  \bs{w}  \big) \big] < 0 \vspace{2mm} \\
								  u_-\big( \la \mid  \bs{w}  \big) \; ,  & \Im\big[ u \big( \la \mid  \bs{w}  \big) \big]  > 0   \ea \right. \quad \e{and} \quad
\psi^{(p)}\big( \la \mid  \bs{w} \big) \; = \; \left\{ \ba{cc} u_+\big( \la \mid  \bs{w}  \big) \; ,  & \Im\big[ u \big( \la \mid  \bs{w}  \big) \big]  > 0 \vspace{2mm} \\
								  u_-\big( \la \mid  \bs{w} \big)  \; , & \Im\big[ u \big( \la \mid  \bs{w}  \big) \big]  < 0 \ea \right. \;.
\label{definition fcts psi p et h}
\enq
Note that since $u_{\pm}$ are restrictions of $u$ to certain domains. As much as the very definition of $\psi^{(h/p)}$ is concerned, one could have just
used solely $u$. Still, the use of restrictions does allow us to stress the inverse of $u$ that will be arising  when inverting $\psi^{(h/p)}$. Furthermore,
it does provide one with a natural domain, implied by the restriction, where those maps are defined.

\begin{figure}[h]
\begin{center}

 \includegraphics{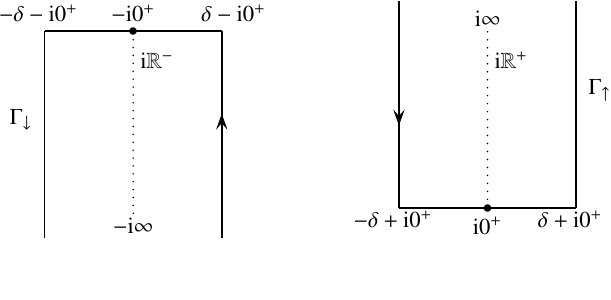}

\caption{Definition of the integration curve $\Ga_{\e{tot}}=\Ga_{\ua} \cup \Ga_{\da}$.}
\label{contour integration Gamma tot}
\end{center}
\end{figure}

  We now introduce the map
\beq
\Psi_n\big(  \bs{w}  \big) \; = \; \Big( \psi^{(p)}\big( y_1 \mid  \bs{w} \big), \dots, \psi^{(p)}\big( y_{n-1} \mid  \bs{w} \big) ,
\psi^{(h)}\big( x_1 \mid  \bs{w}  \big), \dots, \psi^{(h)}\big( x_{n} \mid  \bs{w} \big) \Big) \;.
\label{definition fct Psin}
\enq
Observe that
\beq
\Psi_n \; = \; \Psi_n^{(0)} \, + \, \de\Psi_n \quad \e{with} \quad \de \Psi_n \; = \; \e{O}(T) \;,
\enq
and where
\beq
\Psi_n^{(0)}\big(  \bs{w}  \big) \; = \; \Big( \veps\big( y_1 \big), \dots, \veps\big( y_{n-1} \big) ,
\veps\big( x_1  \big), \dots, \veps\big( x_{n}  \big) \Big) \;.
\enq
This decomposition ensures that $\Psi_n$ is a biholomorphic map, see \cite{KozFaulmanGohmannLowTNLIERigourousAnalysisForQTMMasslessRegime}.
We now introduce the integration contour
\beq
\msc{C}_{n} \; = \; \Psi^{-1}_n\Big(  \big( \Ga_{\e{tot}}\big)^{n-1} \times  \big( \Ga_{\e{tot}}\big)^{n}  \Big) \;
\qquad \e{in}\, \e{which} \qquad
 \Ga_{\e{tot}} \; = \;  \Ga_{\ua}  \cup  \Ga_{\da} \;.
\label{definition C n Thermal FF series contour}
\enq
The contours $\Ga_{\ua/\da}$ are built from a concatenation of segments, \textit{c.f.} Fig.~\ref{contour integration Gamma tot}
\beqa
\Ga_{\ua} & = & \intff{+\i \infty -\de }{  -\de +\i0^+} \cup \intff{  -\de +\i0^+}{  \de +\i0^+} \cup \intff{  \de +\i0^+}{+\i \infty +\de } \;,  \vspace{2mm} \\
\Ga_{\da} & = & \intff{-\i \infty +\de }{  \de -\i0^+} \cup \intff{  \de -\i0^+}{  -\de -\i0^+} \cup \intff{  -\de -\i0^+}{ - \i \infty +\de }  \;.
\eeqa
The parameter $\de>0$ is assumed to be small enough in $T$, \textit{e.g.} $\de = -C T \ln T$, for some $C>0$.

We have finally introduced enough material to present the thermal form factor series expansion for the transverse thermal
dynamical two-point function \cite{KozGohmannKarbachSuzukiFiniteTDynamicalCorrFcts}:
\beq
\big< \sg_1^- \sg_{m+1}^+(t) \big>_{T} \, = \, (-1)^m \sul{ n \geq 1 }{} \f{ (-1)^n }{ n! (n-1)! } \hspace{-1mm} \Int{ \msc{C}_{n}   }{} \hspace{-1mm}
\f{  \dd^{2n-1}w  }{ (2\i\pi)^{2n-1} }
 \f{   \msc{A}^{-+}\big( \bs{w} \big)  \cdot \ex{ \, \i m \De(\bs{w}) }  }
 { \pl{  a=1 }{n-1} \big[ 1 \, + \, \ex{ -\frac{1}{T}  u( y_a \mid \, \bs{w} )  } \big] \cdot
                        \pl{  a=1 }{n} \big[ 1 \, + \, \ex{ \frac{1}{T}  u( x_a \mid \, \bs{w} ) } \big]  }    \;.
\label{ecriture thermal FF series sg-sg+}
\enq
Above, $\bs{w}$ is parametrised as in \eqref{definition vecteur parametres integres}. The series contains several building blocks:
\begin{itemize}

 \item[i)] $ \msc{A}^{-+}\big( \bs{w} \big)$ corresponds to a product of two so-called off-shell
thermal form factors. We refer to \cite{KozDugaveGohmannThermaxFormFactorsXXZ,KozDugaveGohmannThermaxFormFactorsXXZOffTransverseFunctions}
for a detailed expression for this quantity and its behaviour in the low-$T$ limit. We only mention
the properties essential to our analysis. $ \msc{A}^{-+}\big( \bs{w} \big)  $ is holomorphic in $\bs{w}=\big(\bs{x}_n, \bs{y}_{n-1} \big)^{\op{t}}$
belonging to some open neighbourhood of $\msc{C}_{\veps} \cap \mathbb{H}^{-}$ for the variables $x_a$ and of
$\msc{C}_{\veps} \cap \mathbb{H}^{+}$ for the variables $y_a$ with $\msc{C}_{\veps}$ as defined in \eqref{definition contour C veps}.
Furthermore, $ \msc{A}^{-+}\big( \bs{w} \big)  $ is symmetric with respect to the coordinates of $\bs{x}_n$
and $\bs{y}_{n-1}$ taken separately and vanishes on the diagonal, \textit{viz}. when $x_a=x_b$ or $y_a=y_b$
for some $a\not=b$.

\item[ii)] The \textit{a priori} complex valued phase $\De(\bs{w})$ in the exponent is expressed in terms of the effective momentum \eqref{definition impulsion totale}
and effective energy \eqref{definition energie totale}:
\beq
\De(\bs{w}) \; = \; \mc{P}(\bs{w}) + \f{t}{m} \mc{E}(\bs{w}) \;.
\label{definition phase delta w}
\enq
It follows from the low-$T$ expansions \eqref{ecriture DA P}-\eqref{ecriture DA E}
that $\De(\bs{w}) \, = \, \De_{0}(\bs{w}) \, + \, \e{O}\big( T \big)$,  where the leading term is expressed in terms of
the dressed energy \eqref{definition energie habille et energie nue}
and dressed momentum \eqref{definition dressed momentum} as
\beq
 \De_{0}(\bs{w}) \, = \,  \sul{a=1}{n-1} \Big\{ p(y_a) + \f{t}{m} \veps(y_a) \Big\} \, - \, \sul{a=1}{n} \Big\{ p(x_a) + \f{t}{m} \veps(x_a) \Big\} \;.
\label{definition partie dominante du exposant Delta}
\enq

\item[iii)] $ 1 \, + \, \ex{ -\frac{1}{T}  u( y_a \mid \, \bs{w} )  }$ and $1 \, + \, \ex{ \frac{1}{T}  u( x_a \mid \, \bs{w} ) }$
can be thought of as geometric factors related to the fact that, prior to taking the Trotter limit, the multiple integral could be evaluated by multidimensional residue calculus
what relates it to the infinite Trotter limit of the spectral
decomposition associated with the quantum transfer matrix, see \cite{KozGohmannKarbachSuzukiFiniteTDynamicalCorrFcts} for more details.

\end{itemize}

We would like to stress a technical difference between our present description of the series \eqref{ecriture thermal FF series sg-sg+}
and the one provided in \cite{KozGohmannKarbachSuzukiFiniteTDynamicalCorrFcts} . Indeed, in \cite{KozGohmannKarbachSuzukiFiniteTDynamicalCorrFcts}, the series
was written down in terms of abstract contours $\msc{C}_n$ whose existence was simply taken for granted. As a matter of fact, the very construction of the contours
demands an utterly precise control on the structure of the solutions to the higher level Bethe Ansatz equations associated with the quantum transfer matrix.
This was only achieved very recently in \cite{KozFaulmanGohmannLowTNLIERigourousAnalysisForQTMMasslessRegime}, this when $0 \leq \De < 1$ and $T$
is low-enough. This knowledge then allowed us to propose the form of the contours $\msc{C}_n$ as given in \eqref{definition C n Thermal FF series contour}.
For finite $n$, this is obviously the only contour possible, up to homotopy transformations. We believe that it remains adapted even when $n \gg 1$
and $T$ is fixed but small.
Indeed, we trust that \eqref{ecriture thermal FF series sg-sg+} is convergent. In such a case, the behaviour of the coefficients of the series at $\infty$
should not influence the value of the sum taken as a whole.

\section{The large-$(m,t)$ space-like regime of thermal form factor series expansions}
 \label{Section Large m t analysis}

\subsection{A rewriting of the series}
\label{SousSection Reecriture de la series themal FF}

 Observe that the denominator of the integrand in \eqref{ecriture thermal FF series sg-sg+} possesses simple poles at solutions to
\beq
u\big(v \mid \bs{w} \big) \; = \; 2\i\pi T \big(n \, +\, \tfrac{1}{2} \big)
\enq
 with $v=x_a$ or $v=y_b$ for some $a$ or $b$. Note that this a coupled equation owing to the presence of the variable $v$ in the coordinates of $\bs{w}$.
Moreover, the position of the root $v$ does depend implicitly on the other coordinates present in $\bs{w}$.
The purpose of this sub-section is to take explicitly into account the contributions to the thermal form factor series
 \eqref{ecriture thermal FF series sg-sg+}, of the poles close to the endpoints $\pm q$ of the Fermi zone.
This will allow us to explicitly single out the leading contribution to the thermal dynamical transverse two-point function in the low-$T$ regime.

\begin{figure}[h]
\begin{center}

 \includegraphics{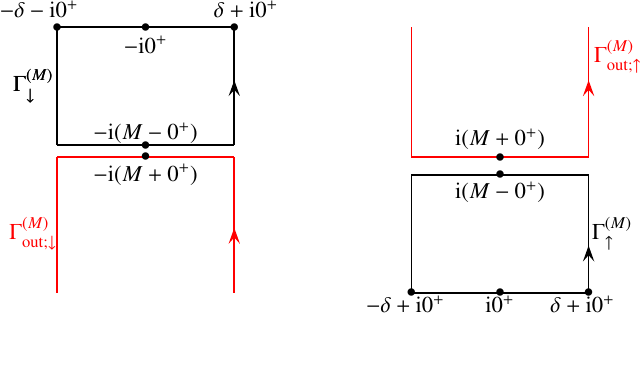}

\caption{Definition of the integration curve $\Ga_{\e{tot}}=\Ga_{\ua} \cup \Ga_{\da}$.}
\label{contour integration Gamma modifiee}
\end{center}
\end{figure}

For that purpose, we introduce a new set of contours depicted in Fig.~\ref{contour integration Gamma modifiee}.  Given $M\in \mathbb{N}^*$, let
\beq
\Ga^{(M)}_{\e{out}} \; = \; \Ga^{(M)}_{\e{out}; \ua } \bigcup \Ga^{(M)}_{\e{out}; \da}
\enq
with
\beqa
\Ga^{(M)}_{\e{out};\ua} & = & \intff{+\i \infty -\de }{  -\de + \i (M +0^+)}
                                \cup \intff{  -\de +\i (M + 0^+) }{  \de + \i (M +0^+) } \cup \intff{  \de + \i (M +0^+) }{+\i \infty +\de }\;,   \nonumber \\
\Ga^{(M)}_{\e{out};\da} & = &  \intff{-\i \infty   +\de }{  \de - \i (M + 0^+) } \cup \intff{  \de - \i (M + 0^+) }{  -\de - \i (M + 0^+) }
\cup \intff{  -\de - \i (M +0^+) }{ - \i \infty +\de } \nonumber
\eeqa
 and further denote
\beqa
\Ga_{\ua}^{(M)} & = & \intff{ \i (M-0^+) -\de  }{  -\de +\i0^+} \cup \intff{  -\de  + \i0^+}{  \de + \i0^+} \nonumber\\
&& \hspace{1cm} \cup \intff{  \de +\i0^+ }{ \i (M - 0^+) + \de  }  \cup \intff{ \i (M -  0^+) + \de  }{ \i (M -0^+) - \de  } \;,  \nonumber  \\
\Ga_{\da}^{(M)} & = & \intff{\de  -\i (M-0^+) }{  \de -\i0^+} \cup \intff{  \de -\i0^+}{  -\de -\i0^+} \nonumber\\
&& \hspace{1cm} \cup \intff{  -\de - \i0^+}{ - \i (M-0^+) - \de }  \cup \intff{ - \i (M-0^+) -\de }{ \de  - \i (M-0^+)  }    \;.
\nonumber
\eeqa

Further, given integers $n_{p/h}^{(\pm)}$, $n_{p/h}^{(M)}$ satisfying
\beq
n_p^{(+)} + n_p^{(-)} + n_p^{(M)} \, = \, n-1  \qquad \e{and} \qquad n_h^{(+)} + n_h^{(-)} + n_h^{(M)} \, = \, n \; ,
\enq
we agree to denote
\beq
\bs{n}_p \, = \,  \big(  n_p^{(-)} , n_p^{(+)} , n_p^{(M)}  \big)^{\op{t}} \qquad \e{as}\; \e{well} \; \e{as} \qquad
\bs{n}_h \, = \,  \big(  n_h^{(-)} , n_h^{(+)} , n_h^{(M)}  \big)^{\op{t}} \;.
\enq
Then we define the associated contour
\beq
\msc{C}_{ \bs{n}_p,\bs{n}_h} \; = \; \Psi_n^{-1}\bigg(  \Big\{ \big\{  \Ga_{\da}^{(M)}  \big\}^{n_p^{(-)}} \times  \big\{  \Ga_{\ua}^{(M)}  \big\}^{n_p^{(+)}}
\times  \big\{  \Ga_{\e{out}}^{(M)}  \big\}^{n_p^{(M)}}  \Big\}
\times 
 \Big\{ \big\{  \Ga_{\ua}^{(M)}  \big\}^{n_h^{(-)}} \times  \big\{  \Ga_{\da}^{(M)}  \big\}^{n_h^{(+)}}  \times  \big\{  \Ga_{\e{out}}^{(M)}  \big\}^{n_h^{(M)}}  \Big\}  \bigg)
\enq
with $\Psi_n$ as defined in \eqref{definition fct Psin}.

Also, for short, we introduce the functions
\beq
\mc{F}(\bs{w}) \; = \;  \f{   \msc{A}^{-+}\big( \bs{w} \big)  \cdot \ex{ \, \i  m \De( \bs{w} ) }  }
 { \pl{  a=1 }{n-1} \Big[ 1 \, + \, \ex{ -\frac{1}{T}  u( y_a \mid \, \bs{w} )  } \Big] \cdot  \pl{  a=1 }{n} \Big[ 1 \, + \, \ex{ \frac{1}{T}  u( x_a \mid \, \bs{w} ) } \Big]  }  \;.
\enq
Then, the symmetry properties of $\mc{F}$ lead to the expansion
\beq
  \Int{ \msc{C}_{n}   }{} \f{  \dd^{2n-1}w }{ (2\i\pi)^{2n-1} } \mc{F}(\bs{w})   \; = \;
\sul{ \substack{  n_p^{(-)} + n_p^{(+)}  \\ + n_p^{(M)} = n-1   } }{}
\sul{ \substack{  n_h^{(-)} + n_h^{(+)}  \\ + n_h^{(M)} = n   } }{}
\f{   n! (n-1)!  }{  \pl{\ups \in \{\pm, M\} }{}  \hspace{-2mm}  n_h^{(\ups)} ! \, n_p^{(\ups)} !  }
  \Int{  \msc{C}_{ \bs{n}_p,\bs{n}_h}   }{}  \pl{\ups \in \{\pm, M\} }{} \bigg\{   \f{  \dd^{  n_p^{(\ups)} } y^{(\ups)} }{ (2\i\pi)^{  n_p^{(\ups)} } }
\cdot \f{  \dd^{  n_h^{(\ups)} } x^{(\ups)} }{ (2\i\pi)^{  n_h^{(\ups)} } }  \bigg\} \cdot \mc{F}(\bs{\varpi}) \; .
\enq
The variable $\bs{w}$ appearing in the \textit{lhs} is as introduced in \eqref{definition vecteur parametres integres}, while
\beq
\bs{\varpi} \; = \; \big(\wt{\bs{x}}_{n} , \wt{\bs{y}}_{n-1} \big)^{ \op{t} }
\enq
appearing in the \textit{rhs} is built from vectors $\wt{\bs{x}}_{n}$ and $\wt{\bs{y}}_{n-1}$ defined by
\beq
\wt{\bs{x}}_{n} \, = \, \Big( \bs{x}_{n_h^{(-)}}^{(-)} ,  \bs{x}_{n_h^{(+)}}^{(+)} , \bs{x}_{n_h^{(M)}}^{(M)}  \Big)^{\op{t}} \qquad \e{and} \qquad
\wt{\bs{y}}_{n-1} \, = \, \Big( \bs{y}_{n_p^{(-)}}^{(-)} ,  \bs{y}_{n_p^{(+)}}^{(+)} , \bs{y}_{n_p^{(M)}}^{(M)}  \Big)^{\op{t}} \; ,
\enq
where
\beq
\bs{x}_{n_h^{(\ups)}}^{(\ups)} \, = \, \Big(x_{1}^{(\ups)}, \dots, x_{n_h^{(\ups)}}^{(\ups)} \Big)^{\op{t}}
\qquad \e{and} \qquad
\bs{y}_{n_p^{(\ups)}}^{(\ups)} \, = \, \Big(y_{1}^{(\ups)}, \dots, y_{n_p^{(\ups)}}^{(\ups)} \Big)^{\op{t}} \; ,
\qquad \e{with} \qquad \ups \in \{\pm, M\}\;.
\enq

The integrals emerging from the preimages of the contours $\Ga_{\ua/\da}^{(M)} $ can be taken explicitly by residue calculations.
For that purpose, we need to introduce several notations.
First of all, consider the vectors
\beq
 \bs{p}_{n_p^{(\sg)}}^{(\sg)} \;= \; \Big( p_1^{(\sg)}, \dots,  p_{n_p^{(\sg)}}^{(\sg)} \Big)^{\op{t}} \qquad \e{and} \qquad
\bs{h}_{n_h^{(\sg)}}^{(\sg)} \;= \; \Big( h_1^{(\sg)}, \dots,  h_{n_h^{(\sg)}}^{(\sg)} \Big)^{\op{t}} \;,
\enq
with $\sg\in \big\{ \pm \big\}$, whose coordinates are integers satisfying  the constraints
\beq
0\leq p_1^{(\sg)} < \cdots < p_{n_p^{(\sg)}}^{(\sg)}\leq M-1\; ,  \qquad  0\leq h_1^{(\sg)} < \cdots < h_{n_h^{(\sg)}}^{(\sg)} \leq M-1 \;.
\label{ecriture choix entiers ordonnés}
\enq
It is further convenient to denote $\bs{p}=\big( \bs{p}_{n_p^{(-)}}^{(-)} , \bs{p}_{n_p^{(+)}}^{(+)}\big)^{\op{t}}$ , $\bs{h}=\big( \bs{h}_{n_h^{(-)}}^{(-)} , \bs{h}_{n_h^{(+)}}^{(+)}\big)^{\op{t}}$.
Next, for fixed vectors $\bs{y}_{n_p^{(M)}}^{(M)}, \bs{x}_{n_h^{(M)}}^{(M)}$, we introduce the unique solution $\op{y}_a^{(\sg)}, \op{x}_a^{(\sg)}$,
belonging to a neighbourhood of $\sg q$,
to the equations
\beq
\ba{ccc} u\big(\op{x}_a^{(\sg)} \mid \bs{\varpi}_{\bs{p},\bs{h}}   \big)
                                \; = \; - \sg 2\i\pi T \big(h_a^{(\sg)}+\tfrac{1}{2} \big) & , &  a=1,\dots, n_h^{(\sg)} \vspace{2mm} \\
u\big(\op{y}_a^{(\sg)} \mid \bs{\varpi}_{\bs{p},\bs{h}}   \big) \; = \;
                \sg 2\i\pi T \big(p_a^{(\sg)}+\tfrac{1}{2} \big)   & , &  a=1,\dots, n_p^{(\sg)}  \ea  \;,
\label{ecriture equation sur les trous et particules effectives}
\enq
where
\beq
\bs{\varpi}_{\bs{p},\bs{h}} \; = \; \Big( \op{x}_{n_h^{(-)}}^{(-)} ,  \op{x}_{n_h^{(+)}}^{(+)} , \bs{x}_{n_h^{(M)}}^{(M)}  ;
\op{y}_{n_p^{(-)}}^{(-)} ,  \op{y}_{n_p^{(+)}}^{(+)} , \bs{y}_{n_p^{(M)}}^{(M)}  \Big)^{\op{t}}
\label{definition point varpi p h}
\enq
and
\beq
 \bs{\op{y}}_{n_p^{(\sg)}}^{(\sg)} \;= \; \Big( \op{y}_1^{(\sg)}, \dots,  \op{y}_{n_p^{(\sg)}}^{(\sg)} \Big)^{\op{t}} \qquad \e{and} \qquad
\bs{\op{x}}_{n_h^{(\sg)}}^{(\sg)} \;= \; \Big( \op{x}_1^{(\sg)}, \dots,  \op{x}_{n_h^{(\sg)}}^{(\sg)} \Big)^{\op{t}} \;.
\label{definition vecteurs y et x sigma}
\enq
The existence and uniqueness of such solutions follows from the work \cite{KozFaulmanGohmannLowTNLIERigourousAnalysisForQTMMasslessRegime}.

We now factorise $\mc{F}\big( \bs{\varpi} \big)$ as
\beq
\mc{F}(   \bs{\varpi} ) \; = \;  \f{ \det \Big[   \op{D}  \Psi_{n}( \bs{\varpi} ) \Big] \cdot   \mc{G}(  \bs{\varpi}  ) }
 { \pl{\sg=\pm}{} \pl{  a=1}{ n_p^{(\sg)}  } \Big[ 1 \, + \, \ex{ -\frac{1}{T}  u( y_a^{(\sg)}  \, \mid  \, \bs{w} )  } \Big] \cdot
 \pl{\sg=\pm}{}  \pl{  a=1}{ n_h^{(\sg)}  }   \Big[ 1 \, + \, \ex{ \frac{1}{T}  u( x_a^{(\sg)}  \, \mid  \, \bs{w} ) } \Big]  }
\enq
in which
\beq
\mc{G}(  \bs{\varpi} ) \; = \; \f{   \msc{A}^{-+}\big(  \bs{\varpi}  \big)  \cdot \ex{ \, \i m \De(  \bs{\varpi} ) }  }{ \det \Big[ \op{D} \Psi_{n}(  \bs{\varpi} ) \Big] }
\cdot \pl{  a=1}{ n_p^{(M)}  } \bigg\{ \f{1}{  1 \, + \, \ex{ -\frac{1}{T}  u( y_a^{(M)}  \mid  \,  \bs{\varpi} )  }  }\bigg\} \cdot
 \pl{  a=1}{ n_h^{(M)}  }\bigg\{ \f{1}{  1 \, + \, \ex{ \frac{1}{T}  u( x_a^{(M)}  \mid \,  \bs{\varpi} )  }  } \bigg\} \;.
\enq

All of the above leads to
\bem
\mc{J}_{ \bs{n}_p,\bs{n}_h} \; = \;
\Int{  \msc{C}_{ \bs{n}_p,\bs{n}_h}   }{}  \pl{\ups \in \{\pm, M\} }{} \bigg\{   \f{  \dd^{  n_p^{(\ups)} } y^{(\ups)} }{ (2\i\pi)^{  n_p^{(\ups)} } }
\cdot \f{  \dd^{  n_h^{(\ups)} } x^{(\ups)} }{ (2\i\pi)^{  n_h^{(\ups)} } }  \bigg\} \cdot \mc{F}(\bs{\varpi})
\; =    \Int{  \Ga_{\e{out}}^{(M)} }{} \hspace{-1mm} \f{ \dd^{n_p^{(M)}} z^{(M)} }{ (2\i\pi)^{n_p^{(M)}} }
  \Int{  \Ga_{\e{out}}^{(M)} }{} \hspace{-1mm} \f{ \dd^{n_h^{(M)}} \mf{z}^{(M)} }{ (2\i\pi)^{n_h^{(M)}} } \\
\times   \Int{  \Ga_{\da}^{(M)} }{} \hspace{-1mm} \f{ \dd^{n_p^{(-)}} z^{(-)} }{ (2\i\pi)^{n_p^{(-)}} }
  \Int{  \Ga_{\ua}^{(M)} }{} \hspace{-1mm} \f{ \dd^{n_h^{(-)}} \mf{z}^{(-)} }{ (2\i\pi)^{n_h^{(-)}} }
  \Int{  \Ga_{\ua}^{(M)} }{} \hspace{-1mm} \f{ \dd^{n_p^{(+)}} z^{(+)} }{ (2\i\pi)^{n_p^{(+)}} }
  \Int{  \Ga_{\da}^{(M)} }{} \hspace{-1mm} \f{ \dd^{n_h^{(+)}} \mf{z}^{(+)} }{ (2\i\pi)^{n_h^{(+)}} }
 \f{ \mc{G}\circ\Psi_n^{-1}\big( \bs{Z}\big) }{  \pl{\sg=\pm}{} \pl{  a=1}{ n_p^{(\sg)}  } \Big[ 1 \, + \, \ex{ -\frac{1}{T}  z^{(\sg)}  } \Big] \cdot
 \pl{\sg=\pm}{}  \pl{  a=1}{ n_h^{(\sg)}  }   \Big[ 1 \, + \, \ex{ \frac{1}{T}  \mf{z}^{(\sg)} } \Big] } \;,
\label{ecriture integrale ctr serie FF pour calcul residus proches}
\end{multline}
where we have introduced the vector
\beq
\bs{Z}\; = \; \Big(  \big( \bs{z}^{(-)}_{n_p^{(-)}},\bs{z}^{(+)}_{n_p^{(+)}}, \bs{z}^{(M)}_{n_p^{(M)}}  \big) ,
                                        \big( \bs{\mf{z}}^{(-)}_{n_p^{(-)}},\bs{\mf{z}}^{(+)}_{n_p^{(+)}}, \bs{\mf{z}}^{(M)}_{n_p^{(M)}}  \big)   \Big)^{\op{t}} \;.
\label{definition vecteur Z}
\enq
One may now take the integrals occurring in the second line of \eqref{ecriture integrale ctr serie FF pour calcul residus proches} in terms of residues
at the simple poles
\beq
z_a^{(\sg)} \, = \, 2\i\pi  \sg \Big( p_a^{(\sg)}+\f{1}{2} \Big) \qquad \e{and} \qquad
\mf{z}_a^{(\sg)} \, = \, - 2\i\pi \sg \Big( h_a^{(\sg)}+\f{1}{2} \Big)
\label{ecriture conditions residus dans variables z}
\enq
%
%
where
\beq
h_a^{(\sg)}, p_a^{(\sg)} \in \intn{0}{M-1} \;.
\enq
Thus, computing the residues will lead, in principle, to a summation over all choices of such integers.
Still, some simplifications are possible. Indeed, $\msc{A}^{-+}\circ \Psi_{n}(\bs{Z})$ vanishes as soon as
two variables of $y$ or two of $x$ type coincide.
It was established in \cite{KozFaulmanGohmannLowTNLIERigourousAnalysisForQTMMasslessRegime} that
solutions to \eqref{ecriture equation sur les trous et particules effectives}
which would be subordinate to some choices of integers such that $h_{a}^{(\sg)}=h_{b}^{(\sg)}$ or $p_{a}^{(\sg)} = p_{b}^{(\sg)}$ for some $a \not= b$
exhibit coinciding coordinates. Thus, one only needs to sum over choices of pairwise distinct integers $h_a^{(\sg)}$ or $p_a^{(\sg)}$
as the other contributions vanish.
Moreover, given a point $\bs{Z}_{\bs{p},\bs{h}}$ computed at the residue coordinates given in \eqref{ecriture conditions residus dans variables z},
\textit{viz}. $\bs{Z}_{\bs{p},\bs{h}}$ is as given in \eqref{definition vecteur Z} in which the coordinates of the
vectors $ \bs{z}^{(\pm)}_{n_p^{(\pm)}}, \bs{\mf{z}}^{(\pm)}_{n_p^{(\pm)}}$ are given by \eqref{ecriture conditions residus dans variables z},
we observe that any permutation of the indices of $h_a^{(\sg)}$ or $p_a^{(\sg)}$ will lead to a permutation
of the associated $"y"$ or $"x"$ type coordinates in $\Psi_{n}^{-1}(\bs{Z}_{\bs{p},\bs{h}})$, hence leaving the value of
$\mc{G}\circ\Psi_n^{-1}\big( \bs{Z}_{\bs{p},\bs{h}}\big)$ invariant, owing to the permutation invariance of $\mc{G}$.
Thus, up to combinatorial factors, one may reduce the summations over
\beq
0 \leq h_1^{(\sg)}  \not=   \cdots \not= h_{n_h^{(\sg)}}^{(\sg)}  \leq M-1 \; , \qquad
0 \leq p_1^{(\sg)}  \not=      \cdots  \not=   p_{n_p^{(\sg)}}^{(\sg)}  \leq M-1 \;,
\enq
to the ordered ones
\beq
0 \leq h_1^{(\sg)} <    \cdots < h_{n_h^{(\sg)}}^{(\sg)}  \leq M-1,  \qquad  0 \leq p_1^{(\sg)} <    \cdots < p_{n_p^{(\sg)}}^{(\sg)}  \leq M-1 \; .
\enq
The above leads to
\beq
\mc{J}_{ \bs{n}_p,\bs{n}_h} \; = \; \Int{  \Ga_{\e{out}}^{(M)} }{} \hspace{-1mm} \f{ \dd^{n_p^{(M)}} z^{(M)} }{ (2\i\pi)^{n_p^{(M)}} }
  \Int{  \Ga_{\e{out}}^{(M)} }{} \hspace{-1mm} \f{ \dd^{n_h^{(M)}} \mf{z}^{(M)} }{ (2\i\pi)^{n_h^{(M)}} }
 \pl{\sg=\pm}{} \Bigg\{ (-T)^{ n_h^{(\sg)} } \cdot  n_h^{(\sg)}!  \hspace{-4mm} \sul{   0 \leq h_1^{(\sg)} <    \cdots < h_{n_h^{(\sg)}}^{(\sg)}    }{ M-1 }
\hspace{-4mm} T^{n_p^{(\sg)}} \cdot n_p^{(\sg)}! \hspace{-4mm} \sul{   0 \leq p_1^{(\sg)} <    \cdots < p_{n_p^{(\sg)}}^{(\sg)}    }{ M-1 } \Bigg\}
\mc{G}\Big( \Psi_{n}^{-1}(\bs{Z}_{\bs{p},\bs{h}}) \Big) \; .
\enq
At this stage one makes the change of integration variables
\beq
\Big(\bs{x}_{n_h^{(M)}}^{(M)}, \bs{y}_{n_p^{(M)}}^{(M)} \Big) \; = \;
\Psi_{n; \e{out}}^{-1}\Big( \bs{z}^{(M)}_{n_p^{(M)}}   , \bs{\mf{z}}^{(M)}_{n_h^{(M)}} \Big)
\enq
where the map $\Psi_{n; \e{out}}$ is defined by
\beq
\Psi_{n; \e{out}}\Big(\bs{x}_{n_h^{(M)}}^{(M)}, \bs{y}_{n_p^{(M)}}^{(M)} \Big) \; = \;
\Big( \psi^{(p)}\big( y_1^{(M)} \mid \bs{\varpi}_{\bs{p},\bs{h}} \big), \dots, \psi^{(p)}\big( y_{ n_p^{(M)}  }^{(M)} \mid  \bs{\varpi}_{\bs{p},\bs{h}} \big) ,
\psi^{(h)}\big( x_1^{(M)} \mid \bs{\varpi}_{\bs{p},\bs{h}}  \big), \dots, \psi^{(h)}\big( x_{ n_h^{(M)}  }^{(M)} \mid \bs{\varpi}_{\bs{p},\bs{h}} \big) \Big) \;.
\enq
We recall that the functions $\psi^{(p/h)}$ have been introduced in \eqref{definition fcts psi p et h} while $\bs{\varpi}_{\bs{p},\bs{h}}$
has been introduced in \eqref{definition point varpi p h}.

Then, upon introducing the integration contour
\beq
\msc{C}_{ n_p^{(M)}\!\! , \, n_h^{(M)}  }  \; = \;
\Psi_{n; \e{out}}^{-1} \bigg(     \big\{  \Ga_{\e{out}}^{(M)}  \big\}^{n_p^{(M)}}   \times   \big\{  \Ga_{\e{out}}^{(M)}  \big\}^{n_h^{(M)}}    \bigg) \;,
\enq
one obtains the below representation for the two-point function:
\bem
\big< \sg_1^- \sg_{m+1}^+(t) \big>_{T} \, = \, \sul{ n \geq 1 }{}  \sul{ \substack{  n_p^{(-)} + n_p^{(+)}  \\ + n_p^{(M)} = n-1   } }{}
\sul{ \substack{  n_h^{(-)} + n_h^{(+)}  \\ + n_h^{(M)} = n   } }{}
\f{1}{  n_p^{(M)} !\,  n_h^{(M)} !}
\pl{\sg=\pm}{} \Bigg\{ (-T)^{ n_h^{(\sg)} }   \hspace{-4mm} \sul{   0 \leq h_1^{(\sg)} <    \cdots < h_{n_h^{(\sg)}}^{(\sg)}    }{ M-1 }
\hspace{-4mm} T^{n_p^{(\sg)}}  \hspace{-4mm} \sul{   0 \leq p_1^{(\sg)} <    \cdots < p_{n_p^{(\sg)}}^{(\sg)}    }{ M-1 } \Bigg\}  \\
 \times \hspace{-2mm}  \Int{  \msc{C}_{ n_p^{(M)}\!\!, \, n_h^{(M)}  }  }{} \hspace{-4mm}
\f{  \dd^{  n_p^{(M)} } y^{(M)}  }{ (2\i\pi)^{  n_p^{(M)} }  } \cdot \f{  \dd^{  n_h^{(M)} } x^{(M)}  }{ (2\i\pi)^{  n_h^{(M)} }   }
\f{ \det\big[ \op{D} \Psi_{n;\e{out}}(  \bs{\varpi}_{\bs{p},\bs{h}} ) \big]    }{ \det \Big[ \op{D} \Psi_{n}(  \bs{\varpi}_{\bs{p},\bs{h}} ) \Big] }
\cdot    \f{  (-1)^{m+n} \msc{A}^{-+}\big(  \bs{\varpi}_{\bs{p},\bs{h}}  \big)  \cdot \ex{ \, \i m \De(  \bs{\varpi}_{\bs{p},\bs{h}} ) }   }
{  \pl{  a=1}{ n_p^{(M)}  } \Big[ 1 \, + \, \ex{ -\frac{1}{T}  u( y_a^{(M)}  \mid  \, \bs{\varpi}_{\bs{p},\bs{h}}  )  } \Big]
 \cdot \pl{  a=1}{ n_h^{(M)}  } \Big[ 1 \, + \, \ex{ \frac{1}{T}  u( x_a^{(M)}  \mid \, \bs{\varpi}_{\bs{p},\bs{h}}   )  } \Big] }    \;.
\label{ecriture serie FF preparee}
\end{multline}
Here, $\bs{\varpi}_{\bs{p},\bs{h}}$ is as defined through \eqref{ecriture equation sur les trous et particules effectives}-\eqref{definition point varpi p h}.

\subsection{The large-$(m,t)$ analysis}
\label{SousSection large m t analysis}

In order to argue, in the low-$T$ limit, the form of the term giving rise to the leading large-$m$ and $t$
asymptotic behaviour of the two-point function with $\big| \tf{t}{m} \big|< c$ and $c>0$ but small, one makes the following assumptions:
\begin{itemize}
 
 \item one may deform $ \msc{C}_{ n_p^{(M)}\!\!, \, n_h^{(M)}  }$ into some $n_p^{(M)} + n_h^{(M)} $ real dimensional sub-manifold
 $ \wt{\msc{C}}_{ n_p^{(M)}\!\!, \, n_h^{(M)}  }$
 of $\Cx^{ n_p^{(M)} + n_h^{(M)}  }$
 which is such that the imaginary parts of the coordinates stay close to zero or $\pm \i\pi /2$ with the exception of those coordinates which have a very large real part, 
 where the given coordinate is supposed to evolve in a region of $\Cx$ where $\Im\big[  p(\om) \big]$ dominates $\Im\big[ \veps (\om) \big]$ and has fixed sign. 
 
 \item Any $"y"$-type coordinate $v$ of a point belonging to this sub-manifold satisfies
 \beq
  C M T  \, < \, \Im\big[  p(v)  \, + \, \tfrac{t}{m} \veps (v) \big] \, < \, \eps
 \label{ecriture borne inf et sup sur contrib dominante particules hors zone M}
 \enq
 for some $C, \eps>0$ $T$-independent. 
 
 \item Any $"x"$-type coordinate $u$ of a point belonging to this sub-manifold satisfies
 \beq
  -C M T  \, > \, \Im\big[  p(u)  \, + \, \tfrac{t}{m} \veps (u) \big] \, > \, - \eps  \;. 
  \label{ecriture borne inf et sup sur contrib dominante trous hors zone M}
 \enq

\end{itemize}

Once these contour-deformation assumptions work, one readily sees that owing to the lower and upper bounds
\eqref{ecriture borne inf et sup sur contrib dominante particules hors zone M}-\eqref{ecriture borne inf et sup sur contrib dominante trous hors zone M}
on the imaginary parts of the leading terms $\De_{0}(\bs{w})$, \textit{c.f.} \eqref{definition partie dominante du exposant Delta},
of the complex valued phase's $\De(\bs{w})$ low-T expansion \cite{KozFaulmanGohmannLowTNLIERigourousAnalysisForQTMMasslessRegime},
the large-$m$ decay issuing from terms in \eqref{ecriture serie FF preparee} which still contain some integrations, \textit{viz}. 
are associated with $n_p^{(M)}\not=0$ or $n_h^{(M)}\not=0$, do give sub-dominant in $m$ contributions, when $M$ is large enough, as compared to
the term which only take into account contributions from the residues of the denominator's poles located closest to $\R$.

We now thus discuss the structure of the latter.
For this purpose, we need to provide the form of the leading low-$T$ expansion of $\De(  \bs{\varpi}_{\bs{p},\bs{h}} )$
in the case when $n_p^{(M)}=n_{h}^{(M)}=0$. The latter was obtained in \cite{KozFaulmanGohmannLowTNLIERigourousAnalysisForQTMMasslessRegime}.
We recall the result here. Let
\beq
0\leq p_1^{(\sg)} < \cdots < p_{n_p^{(\sg)}}^{(\sg)}\qquad \e{and}  \qquad  0\leq h_1^{(\sg)} < \cdots < h_{n_h^{(\sg)}}^{(\sg)}
\enq
be fixed, \textit{viz}. $T$-independent, integers and $\bs{p}, \bs{h}$ the associated vectors of integers
\beq
\left\{ \ba{cc}
\bs{p}\; = \; \big( \bs{p}_{n_p^{(+)}}^{(+)},  \bs{p}_{n_p^{(-)}}^{(-)} \big)^{\op{t}}  \vspace{2mm} \\
\bs{h}\; = \; \big( \bs{h}_{n_h^{(+)}}^{(+)},  \bs{h}_{n_h^{(-)}}^{(-)} \big)^{\op{t}}  \ea \right.
\qquad \e{with} \qquad
 \left\{  \ba{c} \bs{p}_{n_p^{(\sg)}}^{(\sg)} \; = \;  \big( p_1^{(\sg)} \, , \,  \cdots \, , \,  p_{n_p^{(\sg)}}^{(\sg)} \big)^{\op{t}}  \vspace{2mm} \\
\bs{h}_{n_h^{(\sg)}}^{(\sg)} \; = \;  \big( h_1^{(\sg)} \, , \,  \cdots \, , \,  h_{n_h^{(\sg)}}^{(\sg)} \big)^{\op{t}}  \ea \right.  \; .
\enq
Further, let
\beq
\bs{\varpi}_{\bs{p},\bs{h}} \; = \; \Big( \op{x}_{n_h^{(-)}}^{(-)} ,  \op{x}_{n_h^{(+)}}^{(+)}  ;
\op{y}_{n_p^{(-)}}^{(-)} ,  \op{y}_{n_p^{(+)}}^{(+)}  \Big)^{\op{t}}
\enq
be the vector, built out of \eqref{definition vecteurs y et x sigma}, whose coordinates solve
\eqref{ecriture equation sur les trous et particules effectives}. We stress that this time the roots
$\bs{y}^{(M)}_{n_p^{(M)}}, \bs{x}^{(M)}_{n_h^{(M)}}$ are absent. Then, it holds that
\beqa
\mc{P}\big( \bs{\varpi}_{\bs{p},\bs{h}} \big) & = & (\ell^{(-)}  + \ell^{(+)} + 1) p(q) \, + \,
T \mc{P}_0\big( \bs{p},\bs{h} \big)  \; + \; \e{O}\big( T^2 \big) \;,  \\
\mc{E}\big( \bs{\varpi}_{\bs{p},\bs{h}} \big) & = & T \mc{E}_0\big( \bs{p},\bs{h} \big)  \; + \; \e{O}\big( T^2 \big) \;,
\eeqa
in which $\ell^{(\sg)}=\sg \big( n_p^{(\sg)}-n_h^{(\sg)} \big)$ and
\beqa
\mc{P}_0\big( \bs{p},\bs{h} \big)  & = &  \f{2\i\pi}{\op{v}_F}
        \Bigg\{ Z^{2}(q) \big(  \ell^{(-)} \big)^2 + \f{1}{4   Z^{2}(q)  }
            + \sul{\sg=\pm }{} n_p^{(\sg)} n_h^{(\sg)}  \\
 & &   \hspace{5mm} \, + \, \sul{\sg=\pm}{}
\bigg[  \sul{a=1}{ n_p^{(\sg)} } \Big( p_{a}^{(\sg)} \, - \,(a-1) \Big) \; + \;
\sul{a=1}{ n_h^{(\sg)} } \Big( h_{a}^{( \sg)} \, - \, (a-1) \Big)       \bigg]
	       \Bigg\}  \; , \\
\mc{E}_0\big( \bs{p},\bs{h} \big)  & = &  2\i\pi
        \bigg\{ -  \ell^{(-)}   + \sul{\sg=\pm }{} \sg n_p^{(\sg)} n_h^{(\sg)}  \bigg\} \; + \;
    \sul{\sg=\pm}{} \sg
\bigg[  \sul{a=1}{ n_p^{(\sg)} } \Big( p_{a}^{(\sg)}\, - \,(a-1) \Big) \; + \;
\sul{a=1}{ n_h^{(\sg)} } \Big( h_{a}^{( \sg)} \, - \,(a-1) \Big)    \bigg]    \Bigg\} \;.
\eeqa
There $\op{v}_F$ stands for the Fermi velocity, \textit{viz}.
\beq
\op{v}_F= \tf{ \veps^{\prime}(q) }{  p^{\prime}(q) } \; .
\label{definition Fermi velocity}
\enq
The above entails that, to the leading order in $T$,
\beq
\De\big( \bs{\varpi}_{\bs{p},\bs{h}} \big) \, = \,  (\ell^{(-)}  + \ell^{(+)} + 1) p(q) \, + \, T \De_0\big( \bs{\varpi}_{\bs{p},\bs{h}} \big)
 \; + \; \e{O}\big( T^2 \big) \;,
\enq
where $\De_0\big( \bs{\varpi}_{\bs{p},\bs{h}} \big) \, = \,
\De_0^{(1)}\big( \bs{\varpi}_{\bs{p},\bs{h}} \big) + \De_0^{(2)}\big( \bs{\varpi}_{\bs{p},\bs{h}} \big)$ with
\beqa
\De_0^{(1)}\big( \bs{\varpi}_{\bs{p},\bs{h}} \big) & = &  \f{2\i\pi}{\op{v}_F}
        \Bigg\{  \bigg(  Z(q)  \ell^{(-)}  -  \f{\op{v}_F t }{2m  Z(q)  } \bigg)^2
\; + \;   \f{1}{4  Z^2(q) } \bigg( 1  - \Big(   \f{\op{v}_F t }{m   } \Big)^2 \bigg)
            + \sul{\sg=\pm }{} n_p^{(\sg)} n_h^{(\sg)} \Big( 1 + \sg    \f{\op{v}_F t }{m   } \Big)  \Bigg\}  \\
\De_0^{(2)}\big( \bs{\varpi}_{\bs{p},\bs{h}} \big) & = &  \f{2\i\pi}{\op{v}_F}  \sul{\sg=\pm}{}\Big( 1 + \sg    \f{\op{v}_F t }{m   } \Big)
\bigg[  \sul{a=1}{ n_p^{(\sg)} } \Big( p_{a}^{(\sg)} \, - \,(a-1) \Big) \; + \;
\sul{a=1}{ n_h^{(\sg)} } \Big( h_{a}^{( \sg)} \, - \, (a-1) \Big)       \bigg]
	        \;.
\eeqa
At this stage, we simply neglect the contributions of the $\e{O}(T^2)$ remainder and identify the configuration of integers giving rise to the minimal value of
$\Im\big[ \De\big(\bs{\varpi}_{\bs{p},\bs{h}} \big) \big]$ by minimising $\Im\big[ \De_0\big(\bs{\varpi}_{\bs{p},\bs{h}} \big) \big]$.
It is clear that, for
\beq
\big| \tf{\op{v}_F t }{m   } \big|<1 \quad \e{one} \; \e{has} \quad  \Im\big[ \De_0^{(2)}\big(\bs{\varpi}_{\bs{p},\bs{h}} \big) \big] \geq 0
\enq
and that the lower bound is attained for the fully packed configurations
\beq
p_{a}^{(\sg)} \, = \,(a-1)\;, \quad a=1,\dots, n_p^{(\sg)} \qquad \e{and} \qquad
h_{a}^{(\sg)} \, = \,(a-1)\;, \quad a=1,\dots, n_h^{(\sg)}
\label{contrainte fixant p sg et h sg pour minimiser Im Delta 2}
\enq
this irrespectively of the choice of the integers $n_{p/h}^{(\sg)}$. Next, for  $ \big| \tf{\op{v}_F t }{m   } \big|$ small enough,
taken the constraint
\beq
n_p^{(+)}+ n_p^{(-)} + 1 \; = \; n_h^{(+)}+ n_h^{(-)} \; \geq \; 1
\enq
it is clear that the minimal value of $\De_0^{(1)}\big( \bs{\varpi}_{\bs{p},\bs{h}} \big)$ will be attained for
\beq
 \ell^{(-)}=0 \, , \quad \ell^{(+)}= -1 \, , \quad n_{p}^{(-)}=n_p^{(+)}=n_h^{(-)} = 0 \; ,  n_h^{(+)}=1\;.
\enq
The constraints \eqref{contrainte fixant p sg et h sg pour minimiser Im Delta 2}
issuing from minimising $ \Im\big[ \De_0^{(2)}\big(\bs{\varpi}_{\bs{p},\bs{h}} \big) \big]$ then impose that $ h_1^{(+)}=0$. All-in-all, the minimising sequence of integers
for  $\Im\big[ \De_0\big(\bs{\varpi}_{\bs{p},\bs{h}} \big) \big] $ takes the form
\beq
\bs{p}_{\e{min}} \; = \; \emptyset \qquad \e{and} \qquad \bs{h}_{\e{min}} \; = \; \big( \bs{h}_{\e{min}}^{(+)}, \bs{h}_{\e{min}}^{(-)} \big)^{\op{t}}
\quad \e{with} \quad \bs{h}_{\e{min}}^{(+)}\,=\, (0) \quad \e{and} \quad \bs{h}^{(-)}_{\e{min}} \; = \; \emptyset \;.
\enq

All of the above entails that at the minimising configuration $(\bs{p}_{\e{min}},\bs{h}_{\e{min}})$  it holds that
\beq
\De_0\big( \bs{\varpi}_{\bs{p}_{\e{min}},\bs{h}_{\e{min}}} \big) \; = \;  \f{\i\pi}{2\op{v}_F  Z^{2}(q) } \;.
\enq

\subsection{The special case of the free fermion point}
\label{SousSection Free Fermion Results}

It is important for us to stress that one arrives to the very same conclusion when carrying out an approximation free exact analysis
of the free fermion limit, $\De=0$, of the series \eqref{ecriture thermal FF series sg-sg+}, see  \cite{KozGohmannSuzukiXXTransverseSpaceLikeAsymptotics}.
We also remark that the mentioned analysis is valid for any $T$. We summarise the result obtained in \cite{KozGohmannSuzukiXXTransverseSpaceLikeAsymptotics} below.
Given $\De=0$, in the regime $m>4Jt$ it holds that
\beq
\big< \sg_1^- \sg_{m+1}^+(t) \big>_{T} \, = \, (-1)^m \cdot C(T,h) \cdot
\exp\Bigg\{  m \Int{ \mc{C}_{h} }{}  \f{ \dd \la }{2\pi } p_0^{\prime}(\la) \ln \Big| \coth\Big(  \f{\veps_0(\la)}{2T} \Big) \Big| \Bigg\}
\cdot \Big\{  1+ \e{O}\big( m^{-\infty} \big) \Big\} \;.
\enq
The constant $C(T,h)$ is expressed as
\beq
 C(T,h) \; = \; 2T \f{ \Phi(-q) }{ \veps_0^{\prime}(-q) } \exp\Bigg\{ - \Int{ \mc{C}_{h}^{\prime} \subset  \mc{C}_{h} }{} \f{ \dd \la }{ 2\i\pi }
 \Int{   \mc{C}_{h} }{} \f{ \dd \mu }{2\i\pi} \coth^{\prime}(\la-\mu)   \ln  \coth\Big(  \f{\veps_0(\la)}{2T} \Big)  \cdot
 \ln  \coth\Big(  \f{\veps_0(\mu)}{2T} \Big)    \Bigg\} \;.
\enq
There, $\veps_0$ stands for the bare energy introduced in \eqref{definition bare energy} and $p_0$ for the bare momentum introduced in \eqref{definition dressed momentum}.
Both ought to be taken at $\zeta=0$. The integration curve
\beq
\mc{C}_{h}\; = \; \Big\{\R-\i\f{\pi}{2} \Big\}^{\prime} \cup \Big\{-\R   \Big\}^{\prime}
\enq
consists of the two lines $\R-\i\tf{\pi}{2}$ and $\R$, the second one being oriented from $+\infty$ to $-\infty$, with the prescription that the points $-q, q-\i\tf{\pi}{2}$
are inside the domain delimited by these curves while $q, -q - \i \tf{\pi}{2}$ are kept outside.

Finally,
\beq
\Phi(\la) \; = \;    - \f{\i}{2} p_0^{\prime}(\la) \exp\Bigg\{
\Int{   \mc{C}_{h} }{} \f{ \dd \mu }{2 \pi} p_0^{\prime}(\mu)   \ln  \coth\Big(  \f{\veps_0(\la)}{2T} \Big) \cdot \f{ \sinh(\la+\mu + \i \tf{\pi}{4}) }{ \sinh(\la-\mu - \i \tf{\pi}{4}) } \Bigg\}
\enq

The constant $C(T,h)$ can be directly compared with a specific, infinite Trotter number limit, of a product of two
thermal form factors associated with the $\sg^+$ and $\sg^-$ operators. Starting from the finite Trotter number representation for the thermal form factors given in
\cite{KozDugaveGohmannThermaxFormFactorsXXZ,KozDugaveGohmannThermaxFormFactorsXXZOffTransverseFunctions}, one may carry out direct
calculations\symbolfootnote[2]{Here, we stress that the described path leads directly to a comparison with the amplitude $C(T,h)$. If one were to take the $\zeta \tend 0$
limit in the final formulae provided in \cite{KozDugaveGohmannThermaxFormFactorsXXZ,KozDugaveGohmannThermaxFormFactorsXXZOffTransverseFunctions}, then one would find
that $\msc{A}_{\e{dom}}^{-+}$ is expressed in terms of products of Fredholm determinants. While these can be computed and produce ultimately the same answer,
the route to the final result turns out, in fact, more involved that simply starting from scratch and using from the very start the free fermionic structure.},
in the spirit of those explained in \cite{KozDugaveGohmannThermaxFormFactorsXXZ}, so as to obtain that the amplitude $\msc{A}_{\e{dom}}^{-+}$,
expressed as the infinite Trotter number limit of the product of thermal form factors associated with the $\sg^+$ and $\sg^-$ operators and connecting
the dominant Eigenstate with the excited state described by the equations \eqref{equation position dominante trou}-\eqref{ecriture NLIE pour solution sous dominante} at $\De=0$, \textit{viz}.
\beq
u_{\e{dom}} \, = \,  \veps_0 - \i \pi  T \qquad \e{and} \qquad u_{\e{dom}} (x) \; = \; - \i\pi T \quad \e{with} \quad x \in \e{Int}\big( \mc{C}_{h} \big) \;.
\enq
One finds that $x= -q$ and
\beq
\msc{A}_{\e{dom}}^{-+}\; = \; \f{ \veps_0^{\prime}(-q) }{ T } \cdot  C(T,h)  \;.
\enq
In other words, this entails that, at the free fermion point, the leading contribution, in a cone of the space-like regime, is given exactly as stated in
Conjecture \ref{Conjecture pple article}.

\subsection{The conjecture for $|t|\leq  \mf{c} \, |m|$ and any $T$}
\label{SousSection Conjecture for cone in space like regime}

Based on the low-$T$ analysis that we outlined above as well as on the mentioned free fermion results, we formulate the
below conjecture -which we expect to be valid for all ranges of temperatures- on the leading form of the finite temperature large-distance long-time asymptotics of two-point transverse
functions in a cone of the space-like regime.

There exists a constant $c>0$, independent of $T$, such that, when $m \tend +\infty$ and $|t|<c m$,
the transverse thermal dynamical two-point function admits the asymptotic behaviour
\beq
\big< \sg_1^- \sg_{m+1}^+(t) \big>_{T} \, = \, (-1)^m T \f{  \msc{A}^{-+}( \bs{w}_{\e{dom}} ) }{ u^{\prime}( x \mid \bs{w}_{\e{dom}} )  }
\ex{ \i m \De( \bs{w}_{\e{dom}} ) }
\cdot \Big( 1+ \e{O}\big( m^{-\infty} \big) \Big)
\enq
In this expression, $\bs{w}_{\e{dom}}$ is a $1$-dimensional vector, \textit{viz}. there are no $y$ components and there is only one $x$ component:
\beq
\bs{w}_{\e{dom}} \, = \,  \big(  \bs{y}_{\e{dom}} , \bs{x}_{\e{dom}} \big)^{\op{t}} \quad \e{with} \quad\bs{y}_{\e{dom}} \, = \, \emptyset
\;, \; \bs{x}_{\e{dom}} = ( x ) \;.
\enq
Further, the parameter $x$ and the function $u$ solve the coupled system of equations
\beqa
 u( x \mid \bs{w}_{\e{dom}} ) & = & -\i\pi T \;,  \quad \Re(x) >0 \;,  \\
u(\la \mid \bs{w}_{\e{dom}} ) & = & \veps_{0}(\la) \, - \, \i\pi T
\, + \, \i T \th(\la-x) \, - \, T \Int{ \msc{C}_{ u } }{} \dd \mu K(\la-\mu) \msc{L}n\big[1+\ex{-\frac{1}{T}u } \big](\mu \mid \bs{w}_{\e{dom}} ) \;.
\eeqa
Finally, we recall that the complex valued phase $\De$ has been introduced in \eqref{definition phase delta w} and is expressed in terms
of the effective momentum \eqref{definition impulsion totale} and of the effective energy \eqref{definition energie totale}.

\section*{Conclusion}

This work demonstrates the effectiveness of the dynamical thermal form factor series relative to the computation of the large-distance
long-time asymptotic behaviour of transverse two-point functions in some cone of the space-like regime.
These appear to issue from the first term of the series or, equivalently, to be parameterised by the first sub-dominant Eigenvalue of the quantum transfer matrix.
We have checked  our claim against low-$T$ calculations. It is also comforted by our previous analysis of the series
at the free fermion point \cite{KozGohmannSuzukiXXTransverseSpaceLikeAsymptotics}.

\section*{Acknowledgements}
 FG acknowledges financial support by the DFG in the framework
of the research unit FOR 2316. The work of KKK is supported by the
CNRS and by the ERC Project LDRAM: ERC-2019-ADG Project 884584.
The authors thank Andreas Klümper and Junji Suzuki for numerous stimulating discussions
related to the topics tackled in this paper. The authors are indebted to the referee for his constructive comments.

\end{document}